\documentclass[draft]{agujournal2019}
\usepackage{url} 
\usepackage{lineno}
\usepackage[inline]{trackchanges} 
\usepackage{soul}
\usepackage{amsmath}
\draftfalse

\journalname{JGR: Solid Earth}

\begin{document}

%
%


\title{Generalizing scaling laws for mantle convection with mixed heating}

%
%




\authors{Amy L. Ferrick\affil{1}, Jun Korenaga\affil{1}}

\affiliation{1}{Department of Earth and Planetary Sciences, Yale University, New Haven CT 06511}





\correspondingauthor{Amy L. Ferrick}{amy.ferrick@yale.edu}



\begin{keypoints}
\item The boundary layer stability criterion successfully characterizes convection in the mixed heating mode
\item New scaling laws verify the traditional approach for thermal evolution modeling of terrestrial planets
\end{keypoints}

%
%

%
%


\begin{abstract}
Convection in planetary mantles is in the so-called mixed heating mode; it is driven by heating from below, due to a hotter core, as well as heating from within, due to radiogenic heating and secular cooling. Thus, in order to model the thermal evolution of terrestrial planets, we require the parameterization of heat flux for mixed heated convection in particular. However, deriving such a parameterization from basic principles is an elusive task. While scaling laws for purely internal heating and purely basal heating have been successfully determined using the idea that thermal boundary layers are marginally stable, recent theoretical analyses have questioned the applicability of this idea to convection in the mixed heating mode. Here, we present a scaling approach that is rooted in the physics of convection, including the boundary layer stability criterion. We show that, as long as interactions between thermal boundary layers are properly accounted for, this criterion succeeds in describing relationships between thermal boundary layer properties for mixed heated convection. The surface heat flux of a convecting fluid is locally determined by the properties of the upper thermal boundary layer, as opposed to globally determined. Our foundational scaling approach can be readily extended to nearly any complexity of convection within planetary mantles.
\end{abstract}

\section*{Plain Language Summary}
Convection occurring in the rocky interiors of terrestrial planets facilitates their cooling over time. Convection in planetary mantles--the rocky layer bounded by a thin crust and a metallic core--is driven by heat generated within the mantle and heat provided from the underlying metallic core. This so-called mixed heating mode of convection has been suspected to behave quite differently from convection that is heated either solely from within or solely from below. We derive parameterizations of convective heat transfer in terms of the properties of the convective system. We find that mixed heated convection is governed by the same boundary layer dynamics as the two end-member cases. As a result, we may predict how terrestrial planets cool over time in a manner consistent with the physics of mantle convection.

%
%

%


%
%
%
%

\section{Introduction}
Mantle convection governs the thermal evolution of terrestrial bodies. Modeling planetary thermal evolution is a crucial task, as it allows us to assess a planet's thermal history, for which observations are often sparse, and predict a planet's future thermal state. Thermal evolution modeling may be conducted by running full numerical simulations of mantle convection. However, this approach can be unwieldy due to computational limitations or impossible due to poorly constrained complexities (such as plate tectonics on Earth). As a result, an alternative  modeling approach has often been employed -- namely, parameterized mantle convection, which involves the use of scaling laws for heat transport as a function of internal properties  \cite<e.g.,>[]{stevenson1983magnetism,christensen1985thermal}.

Scaling laws for convection driven by heating from below \cite<Rayleigh-B\'enard convection; e.g.,>[]{turcotte1967finite,parmentier1976studies,jarvis1980convection,christensen1984convection,morris1984boundary,bercovici1992three,solomatov1995scaling,liu2013analyses} and convection driven by heating from within \cite<e.g.,>{parmentier1982thermal,davaille1993transient,grasset1998thermal,parmentier2000three,solomatov2000scaling,korenaga2009scaling,korenaga2010scaling,vilella2017fully} have been extensively studied. The basic principle on which many of these scaling laws rely is the boundary layer stability criterion, which states that a thermal boundary layer (TBL) grows until it becomes unstable and breaks off as an upwelling or downwelling \cite{howard1966convection}. According to Howard's conjecture, TBLs are at a steady state with respect to stability and can be described by a stability criterion (i.e., a critical Rayleigh number). Scaling laws based on the boundary layer stability criterion are highly successful in characterizing convection heated purely from below or purely from within. Parameterizations have been extended to account for many complexities relevant to planetary mantles, including three-dimensional spherical geometry \cite<e.g.,>[]{bercovici1992three,vilella2017fully}, depth-, temperature-, and stress-dependent rheology \cite{christensen1984convection,morris1984boundary,davaille1993transient,solomatov1995scaling,moresi1998mantle,solomatov2000scaling,korenaga2009scaling,korenaga2010scaling}, and compressibility \cite{jarvis1980convection,bercovici1992three,liu2013analyses}.

Although scaling laws for convection with either purely internal heating or purely basal heating have commonly been used for thermal evolution modeling, these scalings are, strictly speaking, inappropriate for this task. Planetary mantles are heated both from below (due to a slowly cooling core) and from within (due to radiogenic heating, secular cooling of the mantle, and, in some cases, tidal heating). Ideally, therefore, thermal evolution modeling should be conducted using scaling laws that are generalized to the mixed heating mode of mantle convection.

Parameterization of mixed heating mantle convection has been elusive. Early numerical studies of mixed heated convection suggested a departure from the behavior predicted by the end-member scaling laws for temperature and/or heat flow \cite{jarvis1982mantle,travis1994convection,puster1995characterization}. Later scaling analyses found that mixed heating scaling laws obtained using the well-founded boundary layer stability criterion are successful for only part of the parameter space investigated\cite{sotin1999three,moore2008heat,vilella2018temperature}. It was suggested that, due to interactions between the top and bottom boundary layers, the boundary layer stability criterion may not apply to the mixed heating mode of mantle convection. If true, such a notion is at odds with the well-founded concept that boundary layers are marginally unstable, the foundational physical principal from which many previous scaling laws are derived.

In this paper, we develop new scaling laws for the mixed heating mode of mantle convection, starting with a handful of basic physical principles. We analyze the physics of interactions between the top and bottom boundary layer, and, as long as these interactions are accounted for, the boundary layer stability criterion is successful in characterizing mixed heated convection. Indeed, our approach can be successfully extended to depth-dependent and temperature-dependent viscosity, as well as spherical geometry. The fact that the boundary layer stability criterion still applies for mixed heating conforms to the notion that convection is driven by marginally stable boundary layers. Additionally, and more importantly, we may continue applying the traditional method of modeling the thermal evolution of planetary mantles. This is because the heat flux through the top and bottom of the mantle is simply governed by the structure of the top and bottom boundary layers, respectively.

The structure of the paper is as follows. We first describe the theoretical formulation of a thermally convecting fluid. Next, we address previous scaling approaches for convection driven by heating from both within and below. We then derive new scaling laws using a set of principles suitable for the mixed heating mode. We then extend the scaling laws to depth-dependent, temperature-dependent viscosity, and spherical geometry. Finally, we discuss the implications of our findings and present an application to the strength of Earth's lithosphere.
\section{Theoretical Formulation}

Thermal convection of an incompressible fluid with internally generated heat is governed by conservation of mass, momentum, and energy, represented by the following respective nondimensional equations:
\begin{linenomath*}
\begin{equation}
    \label{eq:1}
    \nabla \cdot \mathbf{u}^* = 0,
\end{equation}
\end{linenomath*}
\begin{linenomath*}
\begin{equation}
    \label{eq:2}
    - \nabla P^* + \nabla \cdot \left[\eta^* \left(\nabla \mathbf{u}^* + \nabla \mathbf{u}^{*T} \right) \right] + RaT^* \mathbf{e}_z = 0,
\end{equation}
\end{linenomath*}
and
\begin{linenomath*}
\begin{equation}
    \label{eq:3}
    \frac{\partial T^*}{\partial t^*} + \mathbf{u}^* \cdot \nabla T^* = \nabla^2 T^* + H^*.
\end{equation}
\end{linenomath*}
Here, time $t^*$ is normalized by the diffusion timescale $D^2/ \kappa$, where $D$ is the depth of the system and $\kappa$ is thermal diffusivity. Spatial coordinates are normalized by $D$, and thus velocity $\mathbf{u}^*$ is normalized by $\kappa / D$. Viscosity $\eta^*$ is normalized by a reference viscosity $\eta_0$, and dynamic pressure $P^*$ is normalized by $\eta_0 \kappa / D^2$. Temperature $T^*$ is normalized by a reference temperature scale $\Delta T$, $H^*$ is the heat generation rate per unit mass, $H$, normalized by $\rho_0 D^2 / k \Delta T$, where $\rho_0$ is a reference density and $k$ is thermal conductivity. The upward unit vector is represented by $\mathbf{e}_z$. The Rayleigh number, $Ra$, is a nondimensional parameter representing the potential vigor of convection, which is defined as
\begin{linenomath*}
\begin{equation}
    \label{eq:4}
    Ra = \frac{\alpha \rho_0 g \Delta T D^3}{\kappa \eta_0},
\end{equation}
\end{linenomath*}
where $\alpha$ is thermal expansivity and $g$ is acceleration due to gravity. The nondimensional time-averaged heat flux at the top and bottom TBLs, $q^*_t$ and $q^*_b$, respectively, are normalized by $k \Delta T/D$. The top and bottom Nusselt numbers ($Nu_t$ and $Nu_b$, respectively) are defined as the top and bottom heat flux, respectively, normalized by a hypothetical conductive heat flux for a system with the same temperature contrast. For mixed heating in which the nondimensional temperature contrast is fixed at unity, we simply have
\begin{linenomath*}
\begin{subequations}
    \begin{equation}
        \label{eq:5a}
        Nu_t = q^*_t,
    \end{equation}
    \begin{equation}
    \label{eq:5b}
        Nu_b = q^*_b.
    \end{equation}
\end{subequations}
\end{linenomath*}

We develop scaling laws for three different viscosity cases, with corresponding numerical experiments: constant viscosity, depth-dependent viscosity, and temperature-dependent viscosity. For depth-dependent viscosity, we impose a two-layered viscosity structure in which one layer layer has a nondimensional viscosity of 1 and the other layer has a nondimensional viscosity of either 10 or 100. We vary the thickness and position (either at the top or bottom of the domain) of the stiff layer. For temperature-dependent viscosity, we use the following linear-exponential viscosity law:
\begin{linenomath*}
\begin{equation}
    \label{eq:6}
    \eta^*(T^*) = \mathrm{exp} \left[ \theta(1-T^*) \right],
\end{equation}
\end{linenomath*}
where the Frank-Kamenetskii parameter, $\theta$, controls the temperature dependence. The Frank-Kamenetskii parameter is related to activation energy $E$ as
\begin{linenomath*}
\begin{equation}
    \label{eq:6b}
    \theta = \frac{E \Delta T}{R \left(T_S + \Delta T \right)^2},
\end{equation}
\end{linenomath*}
where $R$ is the universal gas constant and $T_S$ is the surface temperature.

All numerical experiments are performed using a finite element code \cite{korenaga2003physics} to solve eqs.~\ref{eq:1}--\ref{eq:3} in a 2-D Cartesian domain with an aspect ratio of 4. The domain is discretized into a grid of $256\times 64$ elements in all experiments except for isoviscous runs with $Ra \geq 10^8$. In order to achieve finer resolution in these high-Ra runs, which have very thin TBLs, the uppermost and lowermost five elements of the $256\times 64$ grid are vertically divided further into four elements each. The top and bottom boundaries are held at $T^*=0$ and $T^*=1$, respectively, and internal heat generation is given by $H^*$, defined above. We employ free-slip boundary conditions. All quantities are measured on a time-averaged and horizontally-averaged temperature profile after the simulation reaches statistical steady-state. We consider a simulation at steady-state when time variations in $Nu_t$ drop below 1\%.

\section{Scaling Laws}
\subsection{Previous Work}
As previously stated, scaling laws for purely internally heated and purely basally heated convection have been successfully derived using the TBL stability criterion. We review these scaling laws here, as successful scaling laws for mixed heating must reduce to the scalings for the end-member cases of purely basal and purely internal heating.

In the case of heating only from below (Rayleigh-B\'enard convection), the heat flux through the top of a 2-D Cartesian domain must be equal to the heat flux through the bottom. As a result, the top and bottom TBLs are symmetric, so that the temperature drop across the top and bottom TBLs ($\Delta T_t$ and $\Delta T_b$, respectively) are both $1/2$:
\begin{linenomath*}
\begin{equation}
    \label{eq:7}
    \Delta T_t = \Delta T_b = 1/2.
\end{equation}
\end{linenomath*}
According to the boundary layer stability criterion, the TBLs are marginally stable, and thus  their local Rayleigh numbers can be described by a critical $Ra$:
\begin{linenomath*}
\begin{equation}
    \label{eq:8}
    Ra_{cr} = Ra\Delta T_t \delta_t^3 = Ra\Delta T_b \delta_b^3,
\end{equation}
\end{linenomath*}
where $\delta_t$ and $\delta_b$ are the thickness of the top and bottom TBL, respectively. The right-hand side of eq.~\ref{eq:8} corresponds to the local Rayleigh number of either the top or bottom TBL. Because the TBLs are conducting by definition, we may write
\begin{linenomath*}
\begin{subequations}
\label{eq:9}
    \begin{equation}
        \label{eq:9a}
        Nu_t = \frac{\Delta T_t}{\delta_t},
    \end{equation}
    \begin{equation}
    \label{eq:9b}
        Nu_b = \frac{\Delta T_b}{\delta_b}.
    \end{equation}
\end{subequations}
\end{linenomath*}
From eqs.~\ref{eq:7}--\ref{eq:9}, we arrive at
\begin{linenomath*}
\begin{equation}
    \label{eq:10}
    Nu_t = \frac{1}{2} \left(\frac{Ra}{Ra_{cr}}  \right)^{1/3}.
\end{equation}
\end{linenomath*}
This is the classic scaling law of the form $Nu_t = \alpha Ra^{\beta}$ for Rayleigh-B\'enard convection, where $\beta \sim 1/3$.

For purely internal heating, there is no bottom TBL, and the top heat flux is simply equal to the internal heating:
\begin{linenomath*}
\begin{equation}
    \label{eq:11}
    q^*_t = \frac{\Delta T_t}{\delta_t} = H^*.
\end{equation}
\end{linenomath*}
However, the Nusselt number is now normalized by the internal temperature (approximately equal to the temperature drop across the top TBL), and this temperature is not known a priori:
\begin{linenomath*}
\begin{equation}
    \label{eq:12}
    Nu_t = \frac{q^*_t}{\Delta T_t}.
\end{equation}
\end{linenomath*}
Eq.~\ref{eq:8} (i.e., the boundary layer stability criterion) still applies, so we can use eqs.~\ref{eq:8}, \ref{eq:11}, and \ref{eq:12} to derive the temperature scale,
\begin{linenomath*}
\begin{equation}
    \label{eq:13}
    \Delta T_t \propto H^{*3/4}Ra^{-1/4},
\end{equation}
\end{linenomath*}
and the Nusselt number,
\begin{linenomath*}
\begin{equation}
    \label{eq:14}
    Nu_t \propto (H^*Ra)^{1/4}.
\end{equation}
\end{linenomath*}

When it comes to convection driven by both heating from within and heating from below, it is not so obvious how to derive scalings for $\Delta T_t$ and $Nu_t$ as a function of $Ra$ and $H^*$ using the boundary layer stability criterion. Previous studies have suggested that the boundary layer stability criterion may not accurately describe the behavior of mixed heated convection because of the effect of upwellings and downwellings that arrive at the opposite TBL, and for part or all of the scaling approaches utilized by these studies, no physical justification is provided. For example, \citeA{sotin1999three} and \citeA{moore2008heat} invoke a scaling for the internal temperature (i.e., $\Delta T_t$) by simply taking a linear combination of the scalings for purely basal heating (eq.~\ref{eq:7}) and purely internal heating (eq.~\ref{eq:13}) to arrive at the form $\Delta T_t \sim 0.5 + \gamma H^{*3/4}Ra^{-1/4}$, where $\gamma$ is some constant. \citeA{sotin1999three} then use the boundary layer stability criterion (eq.~\ref{eq:8}) along with eq.~\ref{eq:9a} to arrive at a scaling for $Nu_t$ of the form $Nu_t \propto Ra^{1/3}\Delta T_t^{4/3}$, using their scaling for $\Delta T_t$. In an alternative approach for $Nu_t$, \citeA{moore2008heat} start with the scaling for purely basal heating and add a term proportional to the internal heating: $Nu_t \propto H^* + Ra^{1/3}$. While these scaling laws are relatively successful, the approach of taking a linear combination of the two end-member cases is not rooted in physical principles. More recently, \citeA{vilella2018temperature} derive a scaling for $Nu_t$ by assuming the sum of functions of each of the two input parameters: $Nu_t = f_1(Ra) + f_2(H^*)$. The authors then use the two end-member cases to solve for $f_1$ and $f_2$. However, the physical motivation behind this particular functionality is unclear. \citeA{vilella2018temperature} then derive a scaling for $\Delta T_t$ by considering the force balance in a marginally stable TBL along with conservation of energy. Their initial scaling, of the form $\Delta T_t \sim H^{*1/4}Nu_t^{1/2}Ra^{-1/4}$, fails in the case of purely basal heating, for which the scaling yields $\Delta T_t = 0$. To remedy this, additional functionalities of $Ra$ are incorporated: $\Delta T_t = f_3(Ra)+f_4(Ra)H^{*1/4}Nu_t^{1/2}Ra^{-1/4}$, where $f_3$ and $f_4$ are determined by considering the end-member cases.

Thus, scaling laws for mixed heated convection have yet to be derived based solely on the physics of convection. While the existing scaling laws discussed above achieve a good fit to numerical experiments, it is unclear why they do so, and it is unclear if such scaling laws are applicable beyond the parameter space investigated by previous studies and beyond isoviscous convection. In the following section, we derive mixed heating scaling laws starting from a set of physical principles.

\begin{table}
\caption{Input parameters and output measurements of numerical simulations for isoviscous convection}
\centering
\begin{tabular}{c c c c c c c c c c c c}
\hline
$Ra$ & $H^*$ & $Nu_t$ & $Nu_b$ & $\Delta T_t^{\mathrm{CR}}$ & $\Delta T_t^{\mathrm{HF}}$ & $\delta_t^{\mathrm{CR}}$ & $\delta_t^{\mathrm{HF}}$ & $\Delta T_b^{\mathrm{CR}}$ & $\Delta T_b^{\mathrm{HF}}$ & $\delta_b^{\mathrm{CR}}$ & $\delta_b^{\mathrm{HF}}$\\
\hline
 $3\times 10^4$&0&   6.89&  6.89&  0.517&  0.500&  0.319&  0.0724& 0.517&  0.500&  0.319&  0.0724\\
 $6\times 10^4$&0&	8.52&	8.52&	0.520&	0.500&	0.253&	0.0583&	0.520&	0.500&	0.253&	0.0583\\
 $8\times 10^4$&0&	7.95&	7.95&	0.539&	0.500&	0.227&	0.0627&	0.539&	0.500&	0.227&	0.0621\\
 $10^5$&        0&	8.51&	8.51&	0.540&	0.501&	0.210&	0.0585&	0.539&	0.499&	0.211&	0.0579\\
 $3\times 10^5$&0&	11.65&	11.66&	0.553&	0.503&	0.145&	0.0423&	0.534&	0.497&	0.147&	0.0412\\
 $6\times 10^5$&0&	14.44&	14.42&	0.553&	0.510&	0.115&	0.0340&	0.526&	0.490&	0.117&	0.0322\\
 $8\times 10^5$&0&	15.77&	15.77&	0.553&	0.511&	0.105&	0.0309&	0.526&	0.489&	0.106&	0.0290\\
 $10^6$&        0&	16.94&	16.96&	0.525&	0.490&	0.099&	0.0267&	0.551&	0.510&	0.097&	0.0284\\
 $10^6$&        1&	17.73&	16.68&	0.570&	0.536&	0.096&	0.0285&	0.508&	0.464&	0.100&	0.0255\\
 $10^6$&        3&	19.54&	16.49&	0.596&	0.586&	0.095&	0.0283&	0.486&	0.414&	0.101&	0.0228\\
 $10^6$&        10&	21.87&	11.84&	0.707&	0.706&	0.090&	0.0308&	0.367&	0.294&	0.111&	0.0230\\
 $3\times10^6$&0&	25.94&	25.84&	0.544&	0.506&	0.068&	0.0170&	0.532&	0.494&	0.068&	0.0165\\
 $3\times10^6$&1&	25.75&	24.52&	0.558&	0.531&	0.067&	0.0181&	0.518&	0.469&	0.069&	0.0164\\
 $3\times10^6$&3&	26.25&	22.93&	0.582&	0.570&	0.066&	0.0192&	0.491&	0.430&	0.070&	0.0161\\
 $3\times10^6$&10&	28.54&	18.45&	0.646&	0.644&	0.064&	0.0201&	0.418&	0.356&	0.074&	0.0168\\
 $3\times10^6$&30&	38.57&	8.71&	0.858&	0.852&	0.058&	0.0197&	0.213&	0.148&	0.093&	0.0150\\
 $10^7$&        1&	35.06&	35.39&	0.554&	0.531&	0.045&	0.0121&	0.516&	0.469&	0.046&	0.0112\\
 $10^7$&        3&	36.29&	33.73&	0.562&	0.547&	0.045&	0.0124&	0.499&	0.453&	0.047&	0.0114\\
 $10^7$&        10&	38.46&	29.06&	0.605&	0.602&	0.044&	0.0128&	0.451&	0.398&	0.049&	0.0115\\
 $10^7$&        30&	47.44&	18.20&	0.745&	0.743&	0.041&	0.0127&	0.320&	0.257&	0.054&	0.0115\\
 $3\times10^7$&  1&	49.41&	48.72&	0.536&	0.520&	0.032&	0.0089&	0.509&	0.480&	0.032&	0.0086\\
 $3\times10^7$&  3&	48.84&	46.58&	0.553&	0.548&	0.032&	0.0089&	0.492&	0.452&	0.033&	0.0085\\
 $3\times10^7$& 10&	52.06&	41.71&	0.574&	0.573&	0.031&	0.0092&	0.465&	0.427&	0.033&	0.0087\\
 $3\times10^7$& 30&	60.74&	31.71&	0.666&	0.667&	0.030&	0.0091&	0.385&	0.333&	0.036&	0.0087\\
 $10^8$&        1&	69.77&	68.43&	0.519&	0.504&	0.022&	0.0069&	0.510&	0.496&	0.022&	0.0069\\
 $10^8$&        3&	70.48&	67.80&	0.534&	0.527&	0.022&	0.0072&	0.495&	0.473&	0.022&	0.0067\\
 $10^8$&        10&	74.34&	63.20&	0.556&	0.558&	0.021&	0.0072&	0.471&	0.442&	0.022&	0.0066\\
 $10^8$&        30&	80.93&	52.20&	0.615&	0.622&	0.021&	0.0074&	0.414&	0.378&	0.023&	0.0069\\
 $3\times10^8$& 1&	95.22&	92.95&	0.520&	0.518&	0.015&	0.0050&	0.495&	0.482&	0.015&	0.0048\\
 $3\times10^8$& 3&	97.72&	91.73&	0.522&	0.522&	0.015&	0.0049&	0.493&	0.478&	0.016&	0.0048\\
 $3\times10^8$&10&  101.51&	88.40&	0.544&	0.550&	0.015&	0.0050&	0.473&	0.450&	0.016&	0.0047\\
 $3\times10^8$&30&  108.59&	78.36&	0.581&	0.590&	0.015&	0.0050&	0.438&	0.410&	0.016&	0.0048\\
 $10^9$&1&	131.92&	  131.97&	0.504&	0.506&	0.010&	0.0034&	0.503&	0.494&	0.011&	0.0033\\
 $10^9$&3&	133.65&	130.05&	0.514&	0.519&	0.010&	0.0034&	0.491&	0.481&	0.011&	0.0033\\
 $10^9$&10&	139.02&	123.26&	0.527&	0.532&	0.010&	0.0034&	0.481&	0.468&	0.011&	0.0034\\
 $10^9$&30&	145.51&	112.55&	0.553&	0.562&	0.010&	0.0034&	0.459&	0.438&	0.011&	0.0034\\
\hline
\end{tabular}
\label{table:1}
\end{table}

\subsection{Scaling laws for mixed heated convection with isoviscous rheology}
\begin{figure}
\centering
  \includegraphics[width=1.1\linewidth]{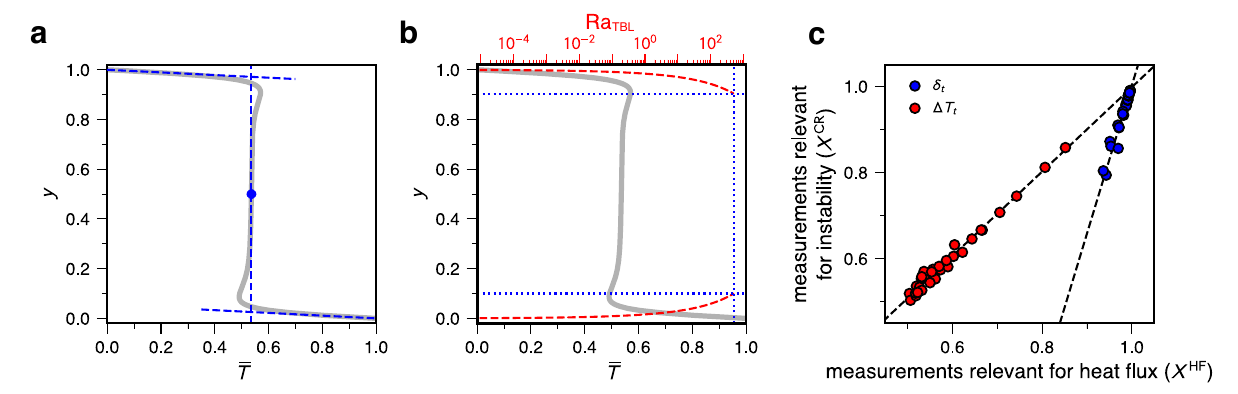}
  \caption{Measurement of thermal boundary layers corresponding to the definitions relevant for (a) heat flux and (b) onset of instability. The TBL structure relevant for heat flux ($\Delta T_t^{\mbox{\scriptsize HF}}$, $\Delta T_b^{\mbox{\scriptsize HF}}$, $\delta_t^{\mbox{\scriptsize HF}}$, and $\delta_b^{\mbox{\scriptsize HF}}$) is determined by where the extension of the temperature gradients at $y=0$ and $y=1$ (sloped dashed blue lines) reach the temperature at the midpoint (vertical dashed blue line). The TBL structure relevant for the onset of instability ($\Delta T_t^{\mbox{\scriptsize CR}}$, $\Delta T_b^{\mbox{\scriptsize CR}}$, $\delta_t^{\mbox{\scriptsize CR}}$, and $\delta_b^{\mbox{\scriptsize CR}}$) is found by calculating the local Rayleigh number of the TBL ($Ra_{\mathrm{TBL}}$) as a function of its hypothetical inner boundary (dashed red line). The inner boundary depth (horizontal dotted blue lines) is then chosen at the depth where $Ra_{\mathrm{TBL}} = Ra_{cr} = 500$ is achieved. In panels (a) and (b), the case with $Ra = 10^6$ and $H^* = 1$ is shown. Panel (c) shows the relationship between the two definitions: $\Delta T_t^{\mbox{\scriptsize CR}}$ vs. $\Delta T_t^{\mbox{\scriptsize HF}}$ is plotted with red circles, and $\delta_t^{\mbox{\scriptsize CR}}$ vs. $\delta_t^{\mbox{\scriptsize HF}}$ is plotted with blue circles. In both cases, the two definitions are related linearly (with lines of best fit plotted as black dashed lines).}
  \label{fig:1}
\end{figure}

We introduce several physical principles regarding a convecting isoviscous fluid, which we use to derive scaling laws. First, when convection is driven by heating from below and within, the heat flux at the top boundary must be the sum of the heat flux at the bottom boundary and the internal heating:
\begin{linenomath*}
\begin{equation}
    \label{eq:15}
    Nu_t = H^* + Nu_b.
\end{equation}
\end{linenomath*}
This relation is based on the conservation of energy. Second, heat flow at the boundaries takes place within conducting thermal boundary layers, such that heat flux is related to the boundary layer structure as
\begin{linenomath*}
\begin{subequations}
\label{eq:16}
    \begin{equation}
        \label{eq:16a}
        Nu_t = \frac{\Delta T_t^{\mbox{\scriptsize HF}}}{\delta_t^{\mbox{\scriptsize HF}}},
    \end{equation}
    \begin{equation}
    \label{eq:16b}
        Nu_b = \frac{\Delta T_b^{\mbox{\scriptsize HF}}}{\delta_b^{\mbox{\scriptsize HF}}}.
    \end{equation}
\end{subequations}
\end{linenomath*}
These equations are the same as eqs.~\ref{eq:9a} and \ref{eq:9b}, but here we make the distinction that the TBL thicknesses and temperature drops are, in this case, those relevant to heat flux (denoted by the superscript ``HF"). This distinction is important because there are several ways of defining the TBLs, and the above relation calls for just one of these definitions. Additionally, when comparing scaling laws with numerical experiments, one must take care to measure TBL properties in a manner consistent with the TBL definition used in the scaling law. For example, $\Delta T_t^{\mbox{\scriptsize HF}}$ and $\delta_t^{\mbox{\scriptsize HF}}$ can be measured by extending the temperature gradient at the upper surface ($y = 1$) until the temperature at the midpoint ($\overline T(y=0.5)$, where $\overline T$ is the time- and horizontally-averaged temperature profile) is reached (Fig.~\ref{fig:1}a). This guarantees that eq.~\ref{eq:16a} is satisfied. Table \ref{table:1} lists the numerical measurements under this definition as well as an alternative definition described below. Note that the structure of the TBL under either definition is hypothetical and not guaranteed to be realized in numerical experiments.

The third governing principle is the boundary layer stability criterion. Previous studies have questioned the applicability of this to mixed heated convection, on account of the interaction of upwellings and downwellings with the opposite TBL \cite{sotin1999three,moore2008heat,vilella2018temperature}. Upon arrival at the opposite TBL, upwellings and downwellings perturb the TBL temperature profile (resulting in the ``overshoot" of TBL temperature past the initial temperature, seen in Fig.~\ref{fig:1}A). Yet, such perturbations alone are not sufficient to prevent the process of TBL growth and break-off of instabilities that ensures the marginal stability of TBLs. For instance, if the temperature perturbations from upwellings and downwellings made a TBL more stable ($Ra_{\mathrm{TBL}} < Ra_{cr}$, where $Ra_{\mathrm{TBL}}$ is the local TBL Rayleigh number), then the TBL would grow conductively until marginal stability is reached. Alternatively, if the temperature perturbations made a TBL more unstable ($Ra_{\mathrm{TBL}} > Ra_{cr}$), then by necessity instabilities would form and break off, returning the TBL to marginal instability. Thus, it is reasonable to assume that the TBLs are still described by marginal stability, and thus the boundary layer stability criterion. A more precise form of the boundary layer stability criterion is given by
\begin{linenomath*}
\begin{equation}
    \label{eq:17}
    Ra_{cr} = Ra \Delta T_t^{\mbox{\scriptsize CR}} \left( \delta_t^{\mbox{\scriptsize CR}} \right)^3 = Ra \Delta T_b^{\mbox{\scriptsize CR}} \left( \delta_b^{\mbox{\scriptsize CR}} \right)^3.
\end{equation}
\end{linenomath*}
Here, the superscript ``CR" refers to a second TBL definition corresponding to the depth at which instability sets in. This guarantees that the local Rayleigh number equals $Ra_{cr}$. We measure $\Delta T_t^{\mbox{\scriptsize CR}}$ and $\delta_t^{\mbox{\scriptsize CR}}$ by assuming some $Ra_{cr}$ and taking the inner boundary of the TBL at the depth where $Ra\Delta T_t^{\mbox{\scriptsize CR}}\left(\delta_t^{\mbox{\scriptsize CR}} \right)^3 = Ra_{cr}$ is achieved (Fig.~\ref{fig:1}b). We choose $Ra_{cr} = 500$, which generally corresponds to the transition from the conducting TBL to the isothermal interior (Fig.~\ref{fig:1}b). The measured values of $\Delta T_t^{\mbox{\scriptsize CR}}$ and $\delta_t^{\mbox{\scriptsize CR}}$ are relatively insensitive to the exact value chosen for $Ra_{cr}$ (see Fig.~\ref{fig:1}b) because $\delta_t^{\mbox{\scriptsize CR}} \propto Ra_{cr}^{1/3}$ (eq.~\ref{eq:17}) and the change in temperature with depth in this region is small. In order to ultimately derive scaling laws, we need to relate the two alternative TBL definitions we have introduced. From our numerical simulations, we find a linear relationship between properties measured by the two different methods (Fig.~\ref{fig:1}c). Thus, we use the following to relate the two TBL definitions:
\begin{linenomath*}
\begin{subequations}
\label{eq:18}
    \begin{equation}
        \label{eq:18a}
        \Delta T_t^{\mbox{\scriptsize CR}} = b\Delta T_t^{\mbox{\scriptsize HF}}, \:\:\: \Delta T_b^{\mbox{\scriptsize CR}} = b\Delta T_b^{\mbox{\scriptsize HF}},
    \end{equation}
    \begin{equation}
    \label{eq:18b}
        \delta_t^{\mbox{\scriptsize CR}} = c\delta_t^{\mbox{\scriptsize HF}}, \:\:\: \delta_b^{\mbox{\scriptsize CR}} = c\delta_b^{\mbox{\scriptsize HF}}.
    \end{equation}
\end{subequations}
\end{linenomath*}
Because $\Delta T_t^{\mbox{\scriptsize HF}}$ and $\Delta T_b^{\mbox{\scriptsize HF}}$ are simply the midpoint temperature and its complement, respectively, and the actual TBL temperature often ``overshoots" this internal temperature, we expect that $b < 1$. On the other hand, by extending the thermal gradient at $y=0$ and $y=1$, we are creating an idealized TBL structure that is thinner than a TBL based on the actual temperature profile. Thus, we expect that $c > 1$.

\begin{figure}
\centering
  \includegraphics[width=1.2\linewidth]{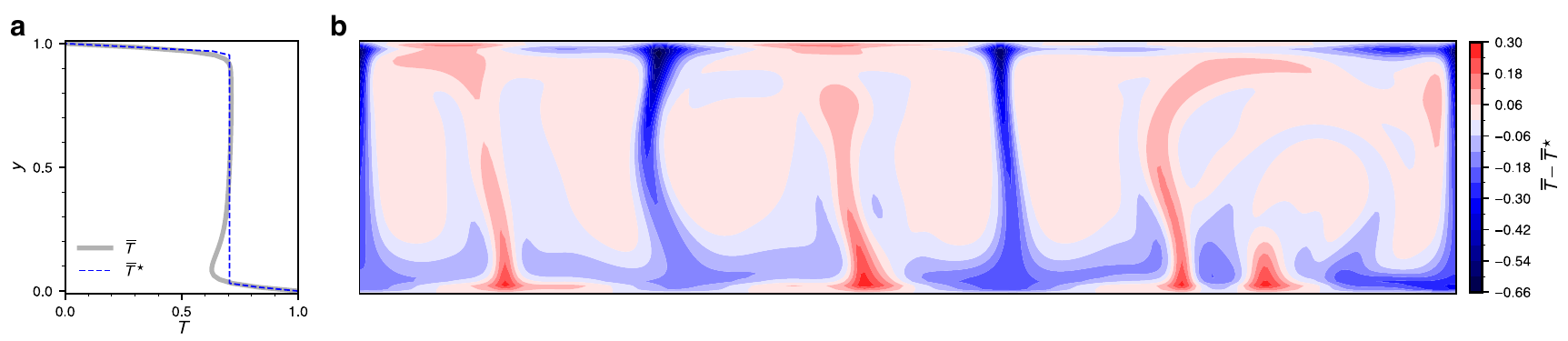}
  \caption{Temperature overshoot due to TBL interaction in isoviscous convection. (a) Time-averaged and horizontally-averaged temperature profile (solid gray curve) and an idealized temperature profile (dashed blue curve) constructed from the internal temperature and the top and bottom heat flux. (b) Temperature anomaly with respect to the idealized temperature profile at a single timestep of the numerical simulation. In both panels, the case with $Ra = 10^6$ and $H^* = 10$ is shown.}
  \label{fig:2}
\end{figure}

The fourth and last constraint is given by the fact that the convecting interior is isothermal, and nearly all of the temperature change occurs in the TBLs. This assumption is valid in the limit of high $Ra$, for which TBLs are well-defined. Under this assumption, we expect that the nondimensional temperature changes across the top TBL, $\Delta T_t^{\mbox{\scriptsize CR}}$, and the the bottom TBL, $\Delta T_b^{\mbox{\scriptsize CR}}$, will sum to 1. However, the temperature at the inner boundary of the top TBL does not equal the temperature at the inner boundary of the bottom TBL; rather, the TBL temperature profiles overshoot the internal temperature, such that the sum of $\Delta T_t^{\mbox{\scriptsize CR}}$ and $\Delta T_b^{\mbox{\scriptsize CR}}$ is greater than 1:
\begin{linenomath*}
\begin{equation}
    \label{eq:19}
    \Delta T_t^{\mbox{\scriptsize CR}} + \Delta T_b^{\mbox{\scriptsize CR}} = 1+\sigma,
\end{equation}
\end{linenomath*}
where $\sigma$ represents the overshoot of $\Delta T_t^{\mbox{\scriptsize CR}} + \Delta T_b^{\mbox{\scriptsize CR}}$ with respect to the net temperature change across the system of 1. In order to derive useful scaling laws, we need to parameterize this overshoot as a function of the dimensionless input parameters. It has been previously speculated that this overshoot is the result of interactions between the boundary layers that perturbs the TBL temperature structure \cite{vilella2018temperature}. To go one step further, we argue that a hot upwelling may not equilibriate with the internal temperature as it rises through the convecting interior, so that it remains hotter than the interior temperature when it reaches the cold upper TBL. Because the upper TBL is conducting, the hot upwelling anomaly comes to rest at the base of the upper TBL, and contributes to a positive thermal anomaly; this is the so-called overshoot. A similar line of reasoning can be made for the effect of cold downwellings on the thermal structure of the lower TBL. The temperature overshoot at the inner boundary of the TBLs can be seen clearly as a deviation of $\overline T(y)$ from an idealized temperature profile constructed from the internal temperature and the temperature gradients at $y=0$ and $y=1$ ($\overline T'$; Fig.~\ref{fig:2}a). As a corollary, in the example shown in Fig.~\ref{fig:2}, most of the overshoot occurs at the bottom TBL because of the large internal heating ratio (defined as $H^*/Nu_t$, or the relative contribution of internal heating to the surface heat flux). In general, however, the total overshoot will be the sum of the overshoot of each TBL with respect to the internal temperature. When we consider the 2-D thermal structure at a single timestep of a numerical simulation, we can clearly see that the deviation from the idealized thermal structure occurs where downwellings (and in some cases, upwellings) are pooling at the base of the opposite TBL (Fig.~\ref{fig:2}b).

We use the following parameterization of the overshoot in our scaling laws:
\begin{linenomath*}
\begin{equation}
    \label{eq:20}
    \sigma = -10.39 Ra^{-1/3}+4.01 Ra^{-0.22}
\end{equation}
\end{linenomath*}
This function, derived in Appendix A, models the measured overshoot well (Fig.~\ref{fig:3}). Its two competing terms are consistent with our intuition. Higher $Ra$ implies faster velocities, and less time for upwellings and downwellings to equilibriate with the internal temperature before reaching the opposite TBL; this contributes to $\sigma$, and is represented by the positive term on the righthand side of eq.~\ref{eq:20}. At the same time, higher $Ra$ implies thinner TBLs, and thus thinner upwellings and downwellings, resulting in a smaller influence on the temperature structure of the opposite TBL; this is represented by the negative term on the lefthand side of eq.~\ref{eq:20}.
\begin{figure}
\centering
  \includegraphics[width=0.4\linewidth]{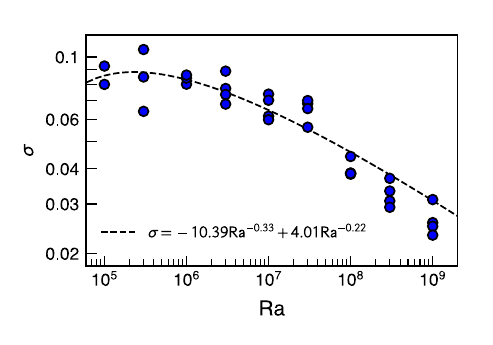}
  \caption{The scaling for the temperature overshoot (dashed black curve) compared to the measured overshoot of the numerical simulations (blue circles). The measured overshoot is taken as $\sigma = \Delta T_t^{\mbox{\scriptsize CR}}+\Delta T_b^{\mbox{\scriptsize CR}} - 1$, consistent with eq.~\ref{eq:19}.}
  \label{fig:3}
\end{figure}

We can solve this system of equations (eqs.~\ref{eq:15}--\ref{eq:20}) for desired properties solely in terms of $Ra$ and $H^*$. First, one may derive the following scaling for $\Delta T_t^{\mbox{\scriptsize HF}}$ in terms of $Ra$, and $H^*$:
\begin{linenomath*}
\begin{equation}
    \label{eq:21}
    (\Delta T_t^{\mbox{\scriptsize HF}})^{4/3} = \left(\frac{1+\sigma}{b} - \Delta T_t^{\mbox{\scriptsize HF}} \right)^{4/3} + \frac{H^*}{c}\left(b \frac{Ra}{Ra_{cr}} \right)^{-1/3}.
\end{equation}
\end{linenomath*}
Whereas $\Delta T_t^{\mbox{\scriptsize HF}}$ cannot be solved for analytically, a numerical solution may be readily obtained for a given pair of $Ra$ and $H^*$. Once $\Delta T_t^{\mbox{\scriptsize HF}}$ is solved for, we can use eqs.~\ref{eq:15}--\ref{eq:20} to obtain other desired parameters. For example, we have
\begin{linenomath*}
\begin{equation}
    \label{eq:22}
    \delta_t^{\mbox{\scriptsize HF}} = \frac{1}{c}\left(b \Delta T_t^{\mbox{\scriptsize HF}} \frac{Ra}{Ra_{cr}} \right)^{1/3},
\end{equation}
\end{linenomath*}
\begin{linenomath*}
\begin{equation}
    \label{eq:23}
    Nu_t = \frac{\Delta T_t^{\mbox{\scriptsize HF}}}{\delta_t^{\mbox{\scriptsize HF}}},
\end{equation}
\end{linenomath*}
\begin{linenomath*}
\begin{equation}
    \label{eq:24}
    \Delta T_t^{\mbox{\scriptsize CR}} = b\Delta T_t^{\mbox{\scriptsize HF}},
\end{equation}
\end{linenomath*}
and
\begin{linenomath*}
\begin{equation}
    \label{eq:25}
    \delta_t^{\mbox{\scriptsize CR}} = c\delta_t^{\mbox{\scriptsize HF}}.
\end{equation}
\end{linenomath*}

\begin{figure}
\centering
  \includegraphics[width=1.2\linewidth]{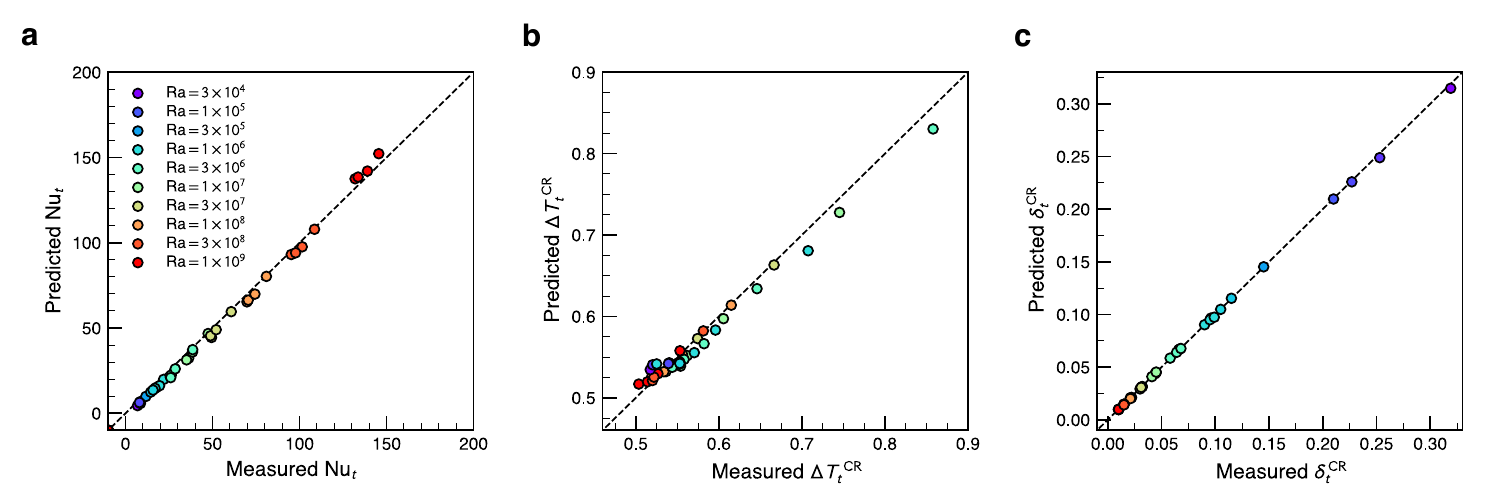}
  \caption{Comparison of the scaling for isoviscous mixed heated convection (eq.~\ref{eq:21}) with numerical experiments. (a) Surface heat flux, (b) top TBL temperature change, (c) top TBL thickness. We include all runs in Table~\ref{table:1}.}
  \label{fig:4}
\end{figure}

We now solve for the best-fit coefficients by fitting the scaling equations to the numerical experiments. We first assume $Ra_{cr} = 500$ as this value was used to measure TBL properties (and thus comparison between measurements and scaling predictions will be justified). For a given pair of $b$ and $c$, the overall misfit is defined as the mean of the normalized squared errors of $Nu_t$, $\Delta T_t^{\mbox{\scriptsize CR}}$, and $\delta_t^{\mbox{\scriptsize CR}}$. The normalized squared error of a property $X$ is $\Sigma (X_{\mathrm{measured}} - X_{\mathrm{predicted}})^2/\Sigma(X_{\mathrm{measured}})^2$, where the sum is over all the numerical runs. The best-fit coefficients are $b = 0.95$ and $c=2.5$, which is close to the values found by comparing the TBL measurements under the two definitions (Fig.~\ref{fig:1}a). The scaling laws predict the results of the numerical experiments very well (Fig.~\ref{fig:4}).

We now verify that the scaling given by eq.~\ref{eq:21} reduces to the well-established scaling laws of the end-member heating modes. This is expected because eq.~\ref{eq:21} is derived using the same physical principles as these end-member scaling laws. In the case of purely basal heating ($H^* = 0$), eq.~\ref{eq:21} yields a $\Delta T_t^{\mbox{\scriptsize HF}}$ that is independent of $Ra$. This is consistent with eq.~\ref{eq:7} and the fact that the TBLs are symmetric in Rayleigh-B\'enard convection regardless of $Ra$. Since $\Delta T_t^{\mbox{\scriptsize HF}}$ is constant, we may use eqs.~\ref{eq:16a} and \ref{eq:17} to arrive at $Nu_t \propto Ra^{1/3}$ which is exactly the classical scaling for Rayleigh-B\'enard convection given by eq.~\ref{eq:10}. In the case of purely internal heating, the temperature scale is initially unknown, and we have $\Delta T_t^{\mbox{\scriptsize HF}}/\delta_t^{\mbox{\scriptsize HF}} = H^*$ and $Nu_t = H^* / \Delta T_t^{\mbox{\scriptsize CR}}$ instead of eq.~\ref{eq:15}. When we further consider the boundary layer stability criterion (eq.~\ref{eq:17}) along with the conversion between TBL definitions (eq.~\ref{eq:18}) we arrive at $Nu_t \propto (H^* Ra)^{1/4}$; this is indeed the traditional scaling given by eq.~\ref{eq:14}.

\begin{figure}
\centering
  \includegraphics[width=0.9\linewidth]{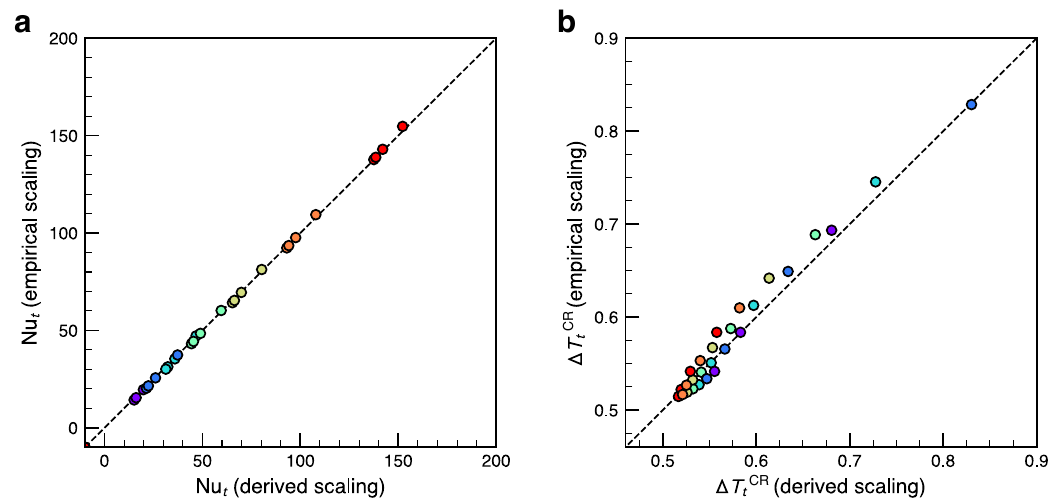}
  \caption{Comparison of the empirical closed-form scalings (eqs.~\ref{eq:26} and \ref{eq:27}) with the exact scalings (eq.~\ref{eq:21} combined with eqs.~\ref{eq:16}--\ref{eq:18}) for (a) surface heat flux and (b) top TBL temperature drop in isoviscous mixed heated convection. Since the derived scalings do not yield closed-form solutions, empirical scalings constructed from the numerical experiments may be useful in the case that numerical solution of the derived scalings is not convenient. Refer to Fig.~\ref{fig:4} for the color scale.}
  \label{fig:5}
\end{figure}

Though eq.~\ref{eq:21} cannot be solved analytically, we may seek ``empirical" scaling laws that express $\Delta T_t^{\mbox{\scriptsize CR}}$ and $Nu_t$ explicitly (i.e., in closed-form) as functions of $Ra$ and $H^*$. Upon inspection of eq.~\ref{eq:21}, we may guess that the numerical measurements will be modeled well by an equation of the form $T_t^{\mbox{\scriptsize HF}} = A'(1+\sigma )/b - \Delta T_t^{\mbox{\scriptsize HF}} + B'\left(H^*/c\right)^{3/4}(bRa/Ra_{cr})^{-1/4}$, where $A'$ and $B'$ are some constants. We can now solve this approximate equation for $\Delta T_t^{\mbox{\scriptsize HF}}$ to get the following relationship:
\begin{linenomath*}
\begin{equation}
    \label{eq:26}
    \Delta T_t^{\mbox{\scriptsize CR}} \approx A + BH^{* 3/4}Ra^{-1/4}.
\end{equation}
\end{linenomath*}
Here, we have converted from $\Delta T_t^{\mbox{\scriptsize HF}}$ to $\Delta T_t^{\mbox{\scriptsize CR}}$ using eq.~\ref{eq:18a} and combined all numerical constants into two coefficients, $A$ and $B$. To complete the empirical scaling law, the combination of $A = 1.038$ and $B = 0.509$ provide the best fit to the numerical simulations. To obtain an empirical scaling for $Nu_t$, we consider eq.~\ref{eq:26} in combination with \ref{eq:16a}, \ref{eq:17}, and \ref{eq:18} to arrive at 
\begin{linenomath*}
\begin{equation}
    \label{eq:27}
    Nu_t \approx CRa^{1/3} + DH^*,
\end{equation}
\end{linenomath*}
where $C$ and $D$ again result from the combination of numerical constants. The best-fit values for these coefficients are  $C = 0.137$ and $D = 0.588$. The empirical closed-form scaling laws given by eqs.~\ref{eq:26} and \ref{eq:27} approximate well our exact scaling given by eq.~\ref{eq:21} (Fig.~\ref{fig:5}). Note that the empirical scaling laws resemble the scaling laws proposed by \citeA{moore2008heat}. While such emprical scaling laws may be reasonable, the exact scaling laws (eqs.~\ref{eq:15}--\ref{eq:19}) are better suited for extension to other rheologies, as they are based on a well-defined set of physical constraints.

\begin{figure}
\centering
  \includegraphics[width=0.75\linewidth]{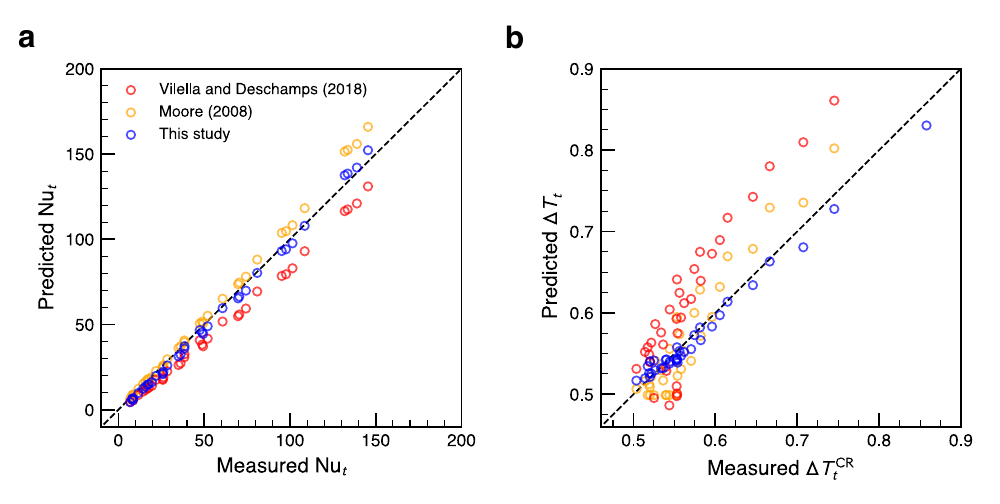}
  \caption{Comparison of proposed scaling laws for (a) heat flux and (b) temperature change across the top TBL. ``This study" refers to eqs.~\ref{eq:21}--\ref{eq:23} assuming $b=0.95$ and $c=2.5$, from which $Nu_t$ and $\Delta T_t^{\mathrm{CR}}$ may be determined. The scalings proposed by \citeA{moore2008heat} are $Nu_t = 1+\frac{1}{2}H^{*}+0.206\left(Ra-658 \right)^{0.318}$ and $\Delta T_t = 0.499+1.33H^{* 3/4}Ra^{-1/4}$. The heat flux and temperature scalings proposed by \citeA{vilella2018temperature} are $Nu_t = \frac{1}{2}\left(H^{*} + H_{cr} \right)$ and $\Delta T_t = \frac{1}{2}\left(\frac{H_{cr}}{2.2} \right)^{3/4}\left(\frac{Ra}{658} \right)^{-1/4}\left(1-\left(\frac{H^{*}}{H_{cr}} \right)^{1/4} \right)+\left(\frac{H^{*}}{2} \right)^{1/4}\left(\frac{Nu_t}{2} \right)^{1/2}\left(\frac{Ra}{658} \right)^{-1/4}$, with $H_{cr}=2+2\left(\frac{Ra}{658}-1 \right)^{1/3}$.}
  \label{fig:6}
\end{figure}

In comparison with previous scaling analyses \cite{moore2008heat,vilella2018temperature}, our scaling law (eq.~\ref{eq:21}, from which $\Delta T_t$ and $Nu_t$ may be determined) better predicts numerical measurements (Fig.~\ref{fig:6}, Table~\ref{table:2}). It should be noted that previous scaling analyses used different methods for measuring TBL properties. These measurements are then used to determine fitting parameters; thus, a comparison of accuracy between different scaling laws is cumbersome and may not be particularly meaningful. Further, the utility of a particular scaling lies not only in its accuracy but also in its capacity for extension to cases that are numerically inaccessible. Because our scaling is derived from physical principles, it may be readily extended beyond two-dimensional isoviscous convection.

\begin{table}
\caption{Accuracy and number of fitting parameters of proposed scaling laws}
\centering
\begin{tabular}{c c c c c}
\hline
 & \multicolumn{2}{c}{Fitting parameters} & \multicolumn{2}{c}{Error$^{a}$}\\
 & $Nu_t$ & $\Delta T_t$ & $Nu_t$ & $\Delta T_t$ \\
\hline
 This study$^{b *}$ & 2 & 2 & 0.0025 & 0.0004\\
 \citeA{moore2008heat} & 2 & 2 & 0.0114 & 0.0033\\
 \citeA{vilella2018temperature}$^{*}$ & 1 & 1 & 0.0255 & 0.0128\\
\hline
\multicolumn{5}{l}{$^a$Normalized squared error as defined in Section 3.2}\\
\multicolumn{5}{l}{$^b$Overshoot scaling parameters were determined prior to fitting $b$ and $c$}\\
\multicolumn{5}{l}{$^*$The scaling laws for $Nu_t$ and $\Delta T_t$ use the same fitting parameters}
\end{tabular}
\label{table:2}
\end{table}

\subsection{Scaling laws for mixed heated convection with depth-dependent viscosity}

\begin{figure}
\centering
  \includegraphics[width=0.75\linewidth]{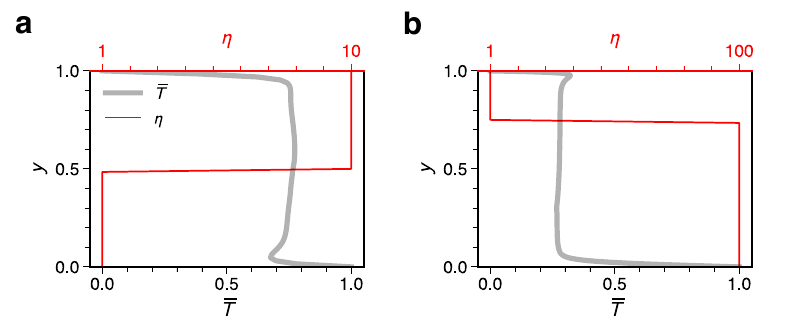}
  \caption{Viscosity profile for two examples of depth-dependent (layered) viscosity convection (red curves), and the corresponding time-averaged and horizontally-averaged temperature profile (gray curves). (a) The case with $Ra = 10^7$, $H^* = 10$, $\eta_{\mathrm{max}} = 10$, and $h = 0.5$, and the stiff layer is overlying the weak layer; (b) the case with $Ra = 3\times 10^8$, $H^* = 10$, $\eta_{\mathrm{max}} = 100$, and $h = 0.75$, and the stiff layer is underlying the weak layer.}
  \label{fig:7}
\end{figure}
We now seek to extend the scaling given by eq.~\ref{eq:21} beyond isoviscous convection, starting with the depth-dependent viscosity described in section 2 (see Table~\ref{table:3} for numerical results). Examples of the viscosity profile and steady-state temperature profile resulting from layered viscosity are shown in Fig.~\ref{fig:7}. Even with depth-dependent viscosity, the boundary layer stability criterion should still apply if we account for TBL viscosity in the local Rayleigh number. We first consider the case in which the high-viscosity layer overlies the low-viscosity layer. In this case, eq.~\ref{eq:17} is modified to
\begin{linenomath*}
\begin{equation}
    \label{eq:28}
    Ra_{cr} = \frac{Ra \Delta T_t^{\mbox{\scriptsize CR}} \left(\delta_t^{\mbox{\scriptsize CR}} \right)^3}{\eta_{\mbox{\scriptsize max}}} = Ra \Delta T_b^{\mbox{\scriptsize CR}} \left(\delta_b^{\mbox{\scriptsize CR}} \right)^3,
\end{equation}
\end{linenomath*}
where $\eta_{\mbox{\scriptsize max}}$ is the viscosity of the stiff layer (either 10 or 100 in our numerical experiments). The bottom TBL has a viscosity of 1 and thus its local $Ra$ is unchanged, but the higher viscosity of the upper TBL must be accounted for. The other assumptions used in the isoviscous scaling remain unaffected, and we arrive at
\begin{linenomath*}
\begin{equation}
    \label{eq:29}
    (\Delta T_t^{\mbox{\scriptsize HF}})^{4/3} = \left(\frac{1+\sigma}{b} - \Delta T_t^{\mbox{\scriptsize HF}} \right)^{4/3} \eta_{\mbox{\scriptsize max}}^{1/3} + \frac{H^*}{c}\left(b \frac{Ra}{Ra_{cr}} \right)^{-1/3} \eta_{\mbox{\scriptsize max}}^{1/3}.
\end{equation}
\end{linenomath*}
In the case of a high-viscosity layer underlying a low-viscosity layer, we follow a similar procedure, this time modifying the local $Ra$ of the lower TBL. The scaling in this case is given by:
\begin{linenomath*}
\begin{equation}
    \label{eq:30}
    (\Delta T_t^{\mbox{\scriptsize HF}})^{4/3} = \left(\frac{1+\sigma}{b} - \Delta T_t^{\mbox{\scriptsize HF}} \right)^{4/3} \eta_{\mbox{\scriptsize max}}^{-1/3} + \frac{H^*}{c}\left(b \frac{Ra}{Ra_{cr}} \right)^{-1/3}.
\end{equation}
\end{linenomath*}
Note that, thus far, the scaling laws for layered viscosity are independent of the thickness of the high-viscosity layer. This is because the lower TBL (or upper TBL, depending on the scenario) is described by $\eta_{\mbox{\scriptsize max}}$ regardless of the thickness of the high-viscosity layer (as long as the TBL is fully contained within the layer).
\begin{table}
\caption{Input parameters and output measurements of numerical simulations with depth-dependent viscosity}
\centering
\begin{tabular}{c c c c c c c c c c}
\hline
$Ra$ & $H^*$ & T/B$^a$ & $\eta_{\mathrm{max}}$ & $h$ & $Nu_t$ & $\Delta T_t^{\mathrm{CR}}$ & $\Delta T_t^{\mathrm{HF}}$ & $\delta_t^{\mathrm{CR}}$ & $\delta_t^{\mathrm{HF}}$ \\
\hline
 $3\times10^6$&3&	T&	10&	0.25&	15.29&	0.745&	0.742&	0.131&	0.0494\\
 $3\times10^6$&3&	T&	10&	0.50&	19.22&	0.676&	0.675&	0.136&	0.0340\\
 $3\times10^6$&3&	T&	10&	0.75&	15.26&	0.733&	0.736&	0.132&	0.0438\\
 $3\times10^6$&3&	T&	100&	0.25&	7.42&	0.891&	0.901&	0.266&	0.1224\\
 $3\times10^6$&3&	T&	100&	0.50&	8.16&	0.899&	0.888&	0.265&	0.1092\\
 $3\times10^6$&3&	T&	100&	0.75&	8.98&	0.899&	0.885&	0.265&	0.0988\\
 $10^7$&10&	B&	10&	0.50&	26.19&	0.486&	0.431&	0.047&	0.0144\\
 $10^7$&10&	B&	100&	0.50&	18.63&	0.426&	0.356&	0.049&	0.0165\\
 $10^7$&10&	T&	10&	0.25&	26.76&	0.757&	0.763&	0.088&	0.0267\\
 $10^7$&10&	T&	10&	0.50&	27.30&	0.755&	0.766&	0.088&	0.0262\\
 $10^7$&10&	T&	10&	0.75&	25.61&	0.775&	0.794&	0.087&	0.0294\\
 $10^7$&10&	T&	100&	0.25&	12.76&	0.990&	0.996&	0.172&	0.0794\\
 $10^7$&10&	T&	100&	0.50&	13.87&	0.976&	0.971&	0.173&	0.0706\\
 $3\times10^7$&3&	B&	10&	0.25&	32.13&	0.410&	0.357&	0.035&	0.0108\\
 $3\times10^7$&10&	B&	100&	0.50&	22.88&	0.374&	0.293&	0.036&	0.0126\\
 $10^8$&3&	B&	10&	0.25&	44.65&	0.389&	0.327&	0.024&	0.0070\\
 $10^8$&10&	B&	100&	0.75&	28.69&	0.318&	0.268&	0.026&	0.0091\\
 $10^8$&30&	T&	10&	0.25&	50.65&	0.822&	0.834&	0.040&	0.0165\\
 $3\times10^8$&3&	T&	10&	0.75&	61.60&	0.667&	0.682&	0.030&	0.0111\\
 $3\times10^8$&10&	B&	100&	0.75&	36.24&	0.316&	0.277&	0.018&	0.0065\\
 $3\times10^8$&30&	B&	10&	0.50&	75.77&	0.484&	0.441&	0.016&	0.0053\\
 $10^9$&3&	B&	100&	0.50&	46.05&	0.265&	0.210&	0.013&	0.0042\\
 $10^9$&10&	T&	10&	0.25&	84.40&	0.678&	0.695&	0.020&	0.0080\\
 $10^9$&30&	B&	10&	0.50&	95.26&	0.445&	0.405&	0.011&	0.0038\\
\hline
\end{tabular}

\footnotesize{$^a$Denotes whether the high-viscosity layer lies at the top (T) or bottom (B) of the domain.}\\
\label{table:3}
\end{table}

The last modification necessary for depth-dependent viscosity is the formulation of the temperature overshoot. The overshoot scaling given by eq.~\ref{eq:20} represents velocity and TBL thicknesses as functions of $Ra$, but for depth-dependent viscosity, $Ra$ (which is defined with a nondimensional viscosity of 1) does not in general predict these convective properties. Therefore, we use a modified Rayleigh number for the overshoot scaling:
\begin{linenomath*}
\begin{equation}
    \label{eq:31}
    \overline{Ra} = \frac{Ra}{\mathrm{exp}\left[\mathrm{log}(\eta_{\mbox{\scriptsize max}})h\right]},
\end{equation}
\end{linenomath*}
where $h$ is the thickness of the stiff layer. We call this the ``log-average $Ra$", because it is normalized by the log-average of the viscosity. The scaling for the temperature overshoot is thus modified to:
\begin{linenomath*}
\begin{equation}
    \label{eq:32}
    \sigma = -10.39 \overline{Ra}^{-1/3}+4.01 \overline{Ra}^{-0.22},
\end{equation}
\end{linenomath*}
Thus, the scaling for depth-dependent viscosity does depend on the thickness of the viscosity layers, although this dependence is a minor one, as $\overline{Ra}$ is not very different from $Ra$, and $\sigma$ itself does not significantly affect the output of the scaling laws.

\begin{figure}
\centering
  \includegraphics[width=1.2\linewidth]{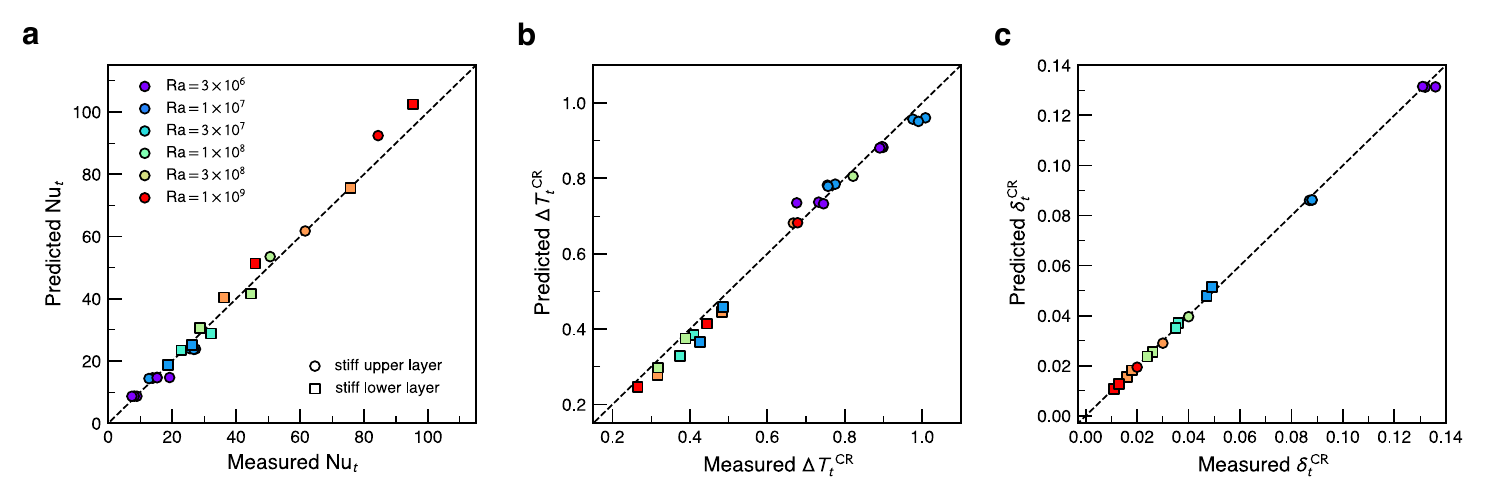}
  \caption{Comparison of numerical simulations with the scaling for mixed heated convection with depth-dependent viscosity (eqs.~\ref{eq:29}--\ref{eq:32}). (a) Surface heat flux, (b) top TBL temperature drop, (c) top TBL thickness.}
  \label{fig:8}
\end{figure}

The validity of eqs.~\ref{eq:29}--\ref{eq:32} can be evaluated by comparing the scaling predictions with numerical experiments. We use the same numerical constants that best fit the isoviscous numerical runs ($Ra_{cr} = 500$, $b = 0.95$, and $c = 2.5$); thus, we are simultaneously evaluating the suitability of these particular numerical constants. The scaling predictions match the measured convective properties remarkably well (Fig.~\ref{fig:8}).

\subsection{Scaling laws for mixed heated convection with temperature-dependent viscosity}
Our next task is to extend the scaling laws to temperature-dependent viscosity given by eq.~\ref{eq:6} (see Table~\ref{table:4} for numerical runs). Under this formulation, there is one additional input parameter: $\theta$, the temperature dependence of viscosity. If $\theta$ is sufficiently large (greater than $\sim$10), then a conducting, immobile lid forms below the surface \cite{solomatov1995scaling}. It is this stagnant lid regime of convection that we seek to derive scaling laws for. This task is more involved than the case of depth-dependent viscosity, but by utilizing scaling arguments developed for purely internally heated stagnant lid convection, we will show that our approach based on boundary layer stability still works.
\begin{table}
\caption{Input parameters and output measurements of numerical simulations with temperature-dependent viscosity}
\centering
\begin{tabular}{c c c c c c c c c c c c}
\hline
$Ra$ & $H^*$ & $\theta$ & $Nu_t$ & $\Delta T_b^{\mathrm{CR}}$ & $\delta_b^{\mathrm{CR}}$ & $\Delta T_L^{v}$ & $D_L^v$ & $\Delta T_{rh}^{\mathrm{CR}}$ & $\delta_{rh}^{CR}$ \\
\hline
 $10^6$&1&	12.0&	2.29&	0.0549&	0.209&	0.189&	0.340&	0.285&	0.129\\
 $10^6$&3&	12.0&	3.04&	0.0036&	0.530&	0.245&	0.312&	0.200&	0.100\\
 $3\times 10^6$&1&	12.0&	2.73&	0.0674&	0.136&	0.225&	0.248&	0.359&	0.126\\
 $3\times 10^6$&3&	12.0&	3.35&	0.0011&	0.547&	0.271&	0.251&	0.256&	0.103\\
 $10^7$&1&	12.0&	3.50&	0.0665&	0.091&	0.289&	0.179&	0.397&	0.107\\
 $10^7$&3&	12.0&	4.30&	0.0278&	0.122&	0.349&	0.170&	0.319&	0.082\\
 $3\times 10^7$&1&	12.0&	4.80&	0.0691&	0.063&	0.398&	0.124&	0.415&	0.079\\
 $3\times 10^7$&3&	12.0&	5.45&	0.0401&	0.075&	0.445&	0.119&	0.375&	0.070\\
 $10^8$&3&	12.0&	6.97&	0.0604&	0.044&	0.533&	0.078&	0.402&	0.079\\
 $10^8$&3&	15.0&	5.54&	0.0420&	0.050&	0.646&	0.121&	0.310&	0.122\\
 $10^8$&3&	16.5&	5.09&	0.0355&	0.053&	0.686&	0.141&	0.278&	0.142\\
 $10^8$&3&	18.0&	4.74&	0.0301&	0.055&	0.715&	0.159&	0.255&	0.160\\
 $10^8$&3&	20.0&	4.32&	0.0240&	0.060&	0.748&	0.185&	0.229&	0.186\\
 $10^8$&3&	22.5&	3.98&	0.0184&	0.065&	0.776&	0.212&	0.207&	0.213\\
 $10^8$&6&	12.0&	7.90&	0.0313&	0.055&	0.581&	0.076&	0.381&	0.077\\
 $10^8$&6&	15.0&	6.69&	0.0129&	0.073&	0.675&	0.106&	0.308&	0.107\\
 $10^8$&6&	16.5&	6.25&	0.0044&	0.105&	0.722&	0.123&	0.269&	0.124\\
 $3\times10^8$&3&	15.0&	7.21&	0.0480&	0.033&	0.608&	0.086&	0.339&	0.087\\
 $3\times10^8$&3&	16.5&	6.54&	0.0418&	0.035&	0.650&	0.102&	0.305&	0.103\\
 $3\times10^8$&3&	18.0&	5.95&	0.0365&	0.036&	0.686&	0.119&	0.276&	0.120\\
 $3\times10^8$&3&	20.0&	5.34&	0.0309&	0.038&	0.712&	0.139&	0.256&	0.140\\
 $3\times10^8$&3&	22.5&	4.84&	0.0246&	0.041&	0.748&	0.163&	0.226&	0.164\\
 $3\times10^8$&6&	15.0&	7.93&	0.0270&	0.040&	0.613&	0.080&	0.354&	0.081\\
 $3\times10^8$&6&	16.5&	7.38&	0.0202&	0.044&	0.653&	0.092&	0.322&	0.093\\
 $3\times10^8$&6&	18.0&	6.86&	0.0141&	0.050&	0.693&	0.106&	0.289&	0.107\\
 $3\times10^8$&6&	20.0&	6.39&	0.0073&	0.062&	0.723&	0.012&	0.266&	0.121\\
 $10^9$&3&	18.0&	7.99&	0.0444&	0.023&	0.659&	0.084&	0.293&	0.085\\
 $10^9$&3&	20.0&	6.98&	0.0376&	0.024&	0.688&	0.101&	0.272&	0.102\\
 $10^9$&3&	22.5&	6.13&	0.0307&	0.026&	0.707&	0.119&	0.260&	0.120\\
 $10^9$&6&	15.0&	10.00&	0.0385&	0.024&	0.578&	0.059&	0.376&	0.060\\
 $10^9$&6&	16.5&	9.36&	0.0320&	0.025&	0.630&	0.069&	0.331&	0.070\\
 $10^9$&6&	18.0&	8.36&	0.0266&	0.027&	0.641&	0.079&	0.327&	0.080\\
 $10^9$&6&	20.0&	7.72&	0.0203&	0.030&	0.676&	0.091&	0.298&	0.092\\
 $10^9$&6&	22.5&	7.03&	0.0138&	0.034&	0.717&	0.107&	0.265&	0.108\\
 $10^9$&9&	15.0&	10.98&	0.0238&	0.028&	0.599&	0.056&	0.368&	0.057\\
 $10^9$&9&	16.5&	10.02&	0.0163&	0.032&	0.631&	0.065&	0.345&	0.066\\
 $10^9$&9&	18.0&	9.27&	0.0104&	0.037&	0.660&	0.074&	0.322&	0.075\\
 $10^9$&9&	20.0&	8.72&	0.0040&	0.050&	0.707&	0.085&	0.282&	0.086\\
 $10^9$&12&	15.0&	11.68&	0.0081&	0.040&	0.612&	0.054&	0.370&	0.055\\
 $3\times10^9$&6&	20.0&	9.30&	0.0296&	0.018&	0.661&	0.073&	0.304&	0.074\\
 $3\times10^9$&6&	22.5&	8.44&	0.0232&	0.020&	0.702&	0.086&	0.270&	0.087\\
 $3\times10^9$&9&	15.0&	13.64&	0.0360&	0.017&	0.563&	0.042&	0.392&	0.043\\
 $3\times10^9$&9&	16.5&	12.94&	0.0324&	0.018&	0.696&	0.055&	0.263&	0.056\\
 $3\times10^9$&9&	18.0&	11.37&	0.0256&	0.019&	0.654&	0.059&	0.313&	0.060\\
 $3\times10^9$&9&	20.0&	10.32&	0.0183&	0.021&	0.708&	0.071&	0.266&	0.072\\
 $3\times10^9$&9&	22.5&	9.32&	0.0116&	0.025&	0.732&	0.082&	0.249&	0.083\\
 $3\times10^9$&12&	15.0&	14.56&	0.0276&	0.019&	0.599&	0.042&	0.363&	0.043\\
 $3\times10^9$&12&	16.5&	13.60&	0.0214&	0.020&	0.715&	0.054&	0.253&	0.055\\
 $3\times10^9$&12&	18.0&	11.89&	0.0137&	0.023&	0.668&	0.058&	0.309&	0.059\\
 $3\times10^9$&12&	20.0&	11.22&	0.0060&	0.031&	0.744&	0.069&	0.241&	0.070\\
 $3\times10^9$&15&	15.0&	15.03&	0.0170&	0.022&	0.616&	0.042&	0.355&	0.043\\
 $3\times10^9$&15&	16.5&	14.24&	0.0105&	0.026&	0.704&	0.051&	0.273&	0.052\\
\hline
\end{tabular}
\label{table:4}
\end{table}

The first two constraints used in the isoviscous case are still valid here, which we summarize as:
\begin{linenomath*}
\begin{equation}
    \label{eq:33}
    Nu_t = H^* + \frac{\Delta T_b^{\mbox{\scriptsize HF}}}{\delta_t^{\mbox{\scriptsize HF}}}.
\end{equation}
\end{linenomath*}
The bottom TBL can be defined using the definitions related to heat flux and instability that we are familiar with. Thus, we still have:
\begin{linenomath*}
\begin{subequations}
\label{eq:34}
    \begin{equation}
        \label{eq:34a}
        \Delta T_b^{\mbox{\scriptsize CR}} = b\Delta T_b^{\mbox{\scriptsize HF}},
    \end{equation}
    \begin{equation}
    \label{eq:34b}
        \delta_b^{\mbox{\scriptsize CR}} = c\delta_b^{\mbox{\scriptsize HF}}.
    \end{equation}
\end{subequations}
\end{linenomath*}
As before, we can apply the boundary layer stability criterion to the bottom TBL:
\begin{linenomath*}
\begin{equation}
    \label{eq:35}
    Ra_{cr} = Ra\Delta T_b^{\mbox{\scriptsize CR}} \left(\delta_b^{\mbox{\scriptsize CR}} \right)^3.
\end{equation}
\end{linenomath*}
Here, we assume that the bottom TBL can be described by a nondimensional viscosity of 1. This is because the presence of the stagnant lid leads to internal temperature very close to 1, so that the temperature of the bottom TBL is approximately 1.

\begin{figure}
\centering
  \includegraphics[width=0.4\linewidth]{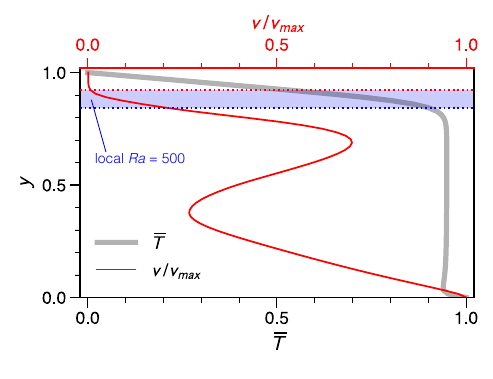}
  \caption{Velocity profile for an example of temperature-dependent viscosity convection (red curve), and the corresponding time-averaged and horizontally-averaged temperature profile (gray curve). The top of the rheological sublayer (dotted red line) is defined where the non-dimensional RMS velocity surpasses 10. The bottom of the rheological sublayer (dotted blue line) is defined by the location where the local $Ra$ of the rheological sublayer becomes $Ra_{cr} = 500$. The example shown is achieved with the following nondimensional parameters: $Ra = 10^8$, $H^*=3$, and $\theta = 12$. The velocity axis is normalized by the maximum RMS velocity; in this case, $v_{\mathrm{max}}\sim1301$.}
  \label{fig:9}
\end{figure}
The top TBL must be treated carefully, as it is comprised of the immobile lid and a rheological sublayer \cite{solomatov2000scaling}. The rheological sublayer conducts heat like the overlying immobile lid but is weak enough to produce downwellings and participate in convection. It is thus reasonable to assume that this rheologial sublayer (but not the entire upper TBL) is marginally unstable and can be characterized by some $Ra_{cr}$. There are then two definitions of the rheological sublayer: one relevant for heat flux, and one relevant for instability. In our numerical experiments, we only measure the sublayer that is relevant for instability (Fig.~\ref{fig:9}). To do so, we first define the base of the immobile lid (and the top of the rheological sublayer) as the depth where the root-mean-square nondimensional velocity exceeds a critical value of 10. We then define the bottom of the rheological sublayer by setting the local $Ra$ equal to $Ra_{cr} = 500$, as we have done previously (Fig.~\ref{fig:9}). We again have the following relationship between the two alternative definitions of the sublayer:
\begin{linenomath*}
\begin{subequations}
\label{eq:36}
    \begin{equation}
        \label{eq:36a}
        \Delta T_{rh}^{\mbox{\scriptsize CR}} = b\Delta T_{rh}^{\mbox{\scriptsize HF}},
    \end{equation}
    \begin{equation}
    \label{eq:36b}
        \delta_{rh}^{\mbox{\scriptsize CR}} = c\delta_{rh}^{\mbox{\scriptsize HF}},
    \end{equation}
\end{subequations}
\end{linenomath*}
where $\Delta T_{rh}$ and $\delta_{rh}$ represent the temperature change across the rheological sublayer and the sublayer thickness, respectively.

Using these definitions of the rheological sublayer, we now turn to establishing some fundamental relations from which we can derive scaling laws. The heat flux through the rheological sublayer must be the sum of the basal heating and the internal heating generated below the immobile lid:
\begin{linenomath*}
\begin{equation}
    \label{eq:37}
    \frac{\Delta T_{rh}^{\mbox{\scriptsize HF}}}{\delta_{rh}^{\mbox{\scriptsize HF}}} = H^*(1-D_L^{v})+ \frac{\Delta T_b^{\mbox{\scriptsize HF}}}{\delta_b^{\mbox{\scriptsize HF}}},
\end{equation}
\end{linenomath*}
where $D_L^{v}$ is the thickness of the immobile lid, defined by the velocity profile as described above (Fig.~\ref{fig:9}). As the rheological sublayer satisfies the boundary layer stability criterion, we may write:
\begin{linenomath*}
\begin{equation}
    \label{eq:38}
    Ra_{cr} = \frac{Ra\Delta T_{rh}^{\mbox{\scriptsize CR}} \left( \delta_{rh}^{\mbox{\scriptsize CR}}\right)^3}{\overline{\eta}},
\end{equation}
\end{linenomath*}
where $\overline{\eta}$ is the log-average of the viscosities at the upper and lower boundary of the rheological sublayer:
\begin{linenomath*}
\begin{equation}
    \label{eq:39}
    \overline{\eta} = \mathrm{exp}\left[\theta \left(1 - \frac{\Delta T_L^{v} + 1 - \Delta T_b^{\mbox{\scriptsize HF}}}{2} \right) \right].
\end{equation}
\end{linenomath*}
Here, $\Delta T_L^{v}$ is the temperature change across the immobile lid as defined by the velocity profile, and we approximate the temperature at the bottom of the rheological sublayer as the temperature at the top of the bottom TBL.

A final constraint on the rheological sublayer is that the temperature difference across it, $\Delta T_{rh}^{\mbox{\scriptsize CR}}$, drives convection and cannot produce a viscosity contrast of more than one order of magnitude, or else some upper portion of the sublayer will be too stiff and incorporate into the immobile lid \cite{solomatov1995scaling,solomatov2000scaling}. This yields the following relationship between $\Delta T_{rh}^{\mbox{\scriptsize CR}}$ and $\theta$:
\begin{linenomath*}
\begin{equation}
    \label{eq:40}
    \Delta T_{rh}^{\mbox{\scriptsize CR}} = a \theta^{-1},
\end{equation}
\end{linenomath*}
where $a$ is an undetermined constant. This scaling of the rheological sublayer was derived by \citeA{solomatov1995scaling} and \citeA{solomatov2000scaling} for purely basally heated convection and purely internally heated convection, respectively, and its applicability to mixed heated convection is reasonable. We find that $a = 4.34$ fits our numerical measurements of $\Delta T_{rh}^{\mbox{\scriptsize CR}}$ best, so we assume this value hereafter. This value of $a$ is somewhat different from that determined by \citeA{solomatov2000scaling}, but this is to be expected because we do not measure the rheological sublayer in the same manner.

A further constraint utilized by \citeA{solomatov2000scaling} is that the immobile lid is characterized by a conductive temperature profile:
\begin{linenomath*}
\begin{equation}
    \label{eq:41}
    \frac{\Delta T_L^{\mbox{\scriptsize HF}}}{D_L^{\mbox{\scriptsize HF}}} = \frac{\Delta T_b^{\mbox{\scriptsize HF}}}{\delta_b^{\mbox{\scriptsize HF}}}+H^* - \frac{1}{2}H^*D_L^{\mbox{\scriptsize HF}}.
\end{equation}
\end{linenomath*}
We have thus far defined the immobile lid using the velocity profile, and this definition may not coincide with where the temperature gradient is conductive. Thus, we have introduced in eq.~\ref{eq:41} a second definition of the lid that is relevant for the conductive temperature gradient (denoted by the superscript ``HF"). There is no reason to assume that these two definitions will be related by the same constants $a$ and $b$ relating the two TBL definitions, as the immobile lid is measured in a different manner. Thus, we introduce
\begin{linenomath*}
\begin{subequations}
\label{eq:42}
    \begin{equation}
        \label{eq:42a}
        \Delta T_{L}^{\mbox{\scriptsize HF}} = d\Delta T_{L}^{v},
    \end{equation}
    \begin{equation}
    \label{eq:42b}
        \delta_{L}^{\mbox{\scriptsize HF}} = e\delta_{L}^{v},
    \end{equation}
\end{subequations}
\end{linenomath*}
where $d$ and $e$ are undetermined constants.

As a final constraint, we may reason that, because the convective interior is relatively isothermal, the temperature changes across the immobile lid, rheological sublayer, and the bottom TBL must sum to 1, the total temperature contrast across the system:
\begin{linenomath*}
\begin{equation}
    \label{eq:43}
    \Delta T_L^{v} + \Delta T_b^{\mbox{\scriptsize CR}} + \Delta T_{rh}^{\mbox{\scriptsize CR}} = 1.
\end{equation}
\end{linenomath*}
Note that we do not include the temperature overshoot $\sigma$ in this constraint. This is because most of the temperature change occurs in the immobile lid, and the temperature change across the sublayer and the bottom TBL are sufficiently small such that boundary layer interactions are negligible.

Scaling laws can finally be obtained by combining eqs.~\ref{eq:33}--\ref{eq:43}. We first derive an equation for $D_L^{v}$ and $\Delta T_b^{\mbox{\scriptsize HF}}$ in terms of the nondimensional input parameters. The equation is quadratic in $D_L^{v}$, and thus has two possible solutions. Upon inspection of measurements of $D_L^{v}$ and $\Delta T_b^{\mbox{\scriptsize CR}}\left(= b\Delta T_b^{\mbox{\scriptsize HF}}\right)$, we determine which of the two solutions is appropriate:
\begin{linenomath*}
\begin{multline}
\label{eq:44}
    D_L^{v} = \frac{1}{e^2}+\frac{c}{e^2 H^*}\left(\Delta T_b^{\mbox{\scriptsize HF}}\right)^{4/3}\left(b\frac{Ra}{Ra_{cr}} \right)^{1/3} \\
    - \frac{d}{e^2 H^*} \left[ \left(-\frac{H^*}{d} - \frac{c}{d}\left(\Delta T_b^{\mbox{\scriptsize HF}}\right)^{4/3}\left(bRa/Ra_{cr} \right)^{1/3} \right)^2 - 2\frac{e^2}{d}H\left(1 -b\Delta T_b^{\mbox{\scriptsize HF}} - a\theta^{-1} \right) \right]^{1/2}.
\end{multline}
\end{linenomath*}
Eqs.~\ref{eq:33}--\ref{eq:43} yield a second equation relating $\Delta T_{b}^{\mbox{\scriptsize HF}}$ and $D_L^{v}$:
\begin{linenomath*}
\begin{multline}
    \label{eq:45}
    c\left(\frac{a\theta^{-1}}{b} \right)^{4/3}\left(b\frac{Ra}{Ra_{cr}} \right)^{1/3}\times \\
    \mathrm{exp}\left[-\frac{\theta}{3}\left(1 - 0.5\left(1+\Delta T_b^{\mbox{\scriptsize HF}}+\frac{H^*}{d}D_L^{v} - \frac{e^2 H^*}{2d}\left(D_L^{v}\right)^2+\frac{c}{d}D_L^{v}\left(\Delta T_b^{\mbox{\scriptsize HF}} \right)^{4/3}\left(b\frac{Ra}{Ra_{cr}} \right)^{1/3} \right) \right) \right] \\
    = H^*\left(1-D_L^{v} \right)+c\left(\Delta T_b^{\mbox{\scriptsize HF}}\right)^{4/3}\left(b\frac{Ra}{Ra_{cr}} \right)^{1/3}.
\end{multline}
\end{linenomath*}
Thus, the two equations can be numerically solved for the two unknowns, $\Delta T_b^{\mbox{\scriptsize HF}}$ and $D_L^{v}$. Because we have already determined that $Ra_{cr} = 500$, $a = 4.34$, $b = 0.95$, and $c = 2.5$, we only need to fit $d$ and $e$ to the numerical measurements. We evaluate the fitness of a given combination of $d$ and $e$ to predict $Nu_t$, $\Delta T_b^{\mbox{\scriptsize CR}}$, and $D_L^{v}$ using the misfit measure introduced in section 3.2. We find that $d = 0.9$ and $e = 0.97$.
\begin{figure}
\centering
  \includegraphics[width=1.2\linewidth]{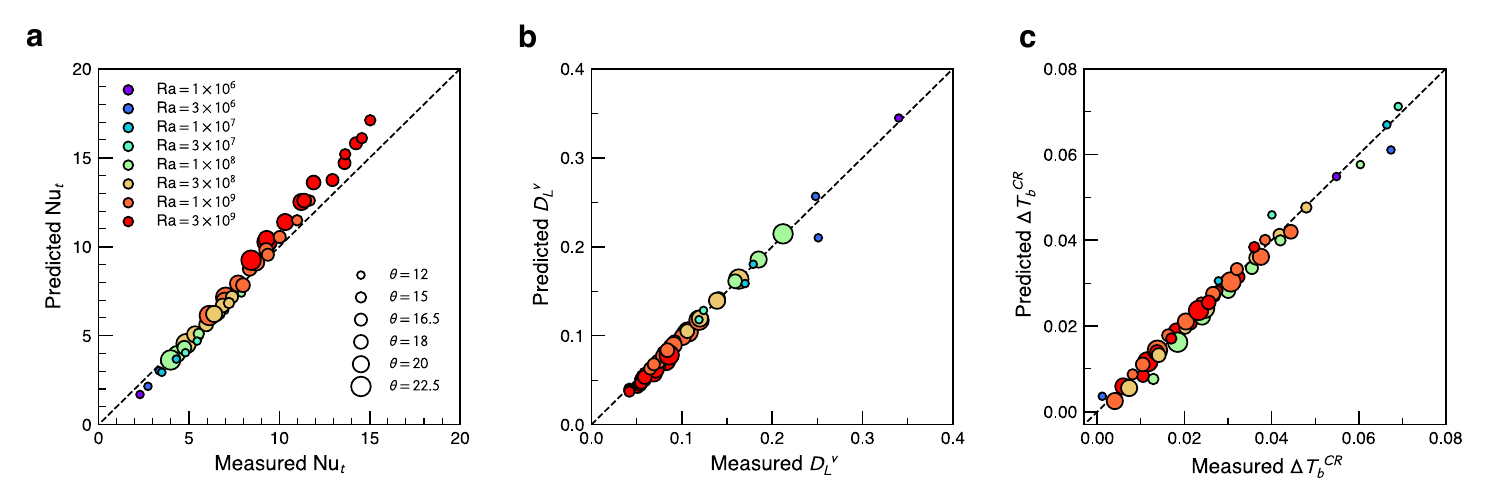}
  \caption{Comparison of numerical simulations with the scaling for mixed heated convection with temperature-dependent viscosity (eqs.~\ref{eq:44} and \ref{eq:45}). (a) Surface heat flux, (b) immobile lid thickness, (c) bottom TBL temperature drop.}
  \label{fig:10}
\end{figure}

The scaling is successful in predicting the measured values of $Nu_t$, $\Delta T_b^{\mbox{\scriptsize CR}}$, and $D_L^{v}$ (Fig.~\ref{fig:10}). We have included several moderate--$Ra$ cases ($Ra < 10^8$), which are characterized by relatively large variations in lid thickness. A few of these cases agree slightly more poorly with the scaling predictions than the high--$Ra$ cases. This is to be expected, as our scaling is based on the assumption of well-defined boundary layers which are ubiquitous only at high $Ra$.

\subsection{Extension to spherical geometry}
To further demonstrate the merit of the boundary layer stability approach, we extend our scaling analysis to spherical geometry in both the isoviscous case and the depth-dependent viscosity case, for which published numerical experiments are available \cite{deschamps2010temperature,o2013comparison,weller2016scaling}. We first consider the isoviscous case.

\begin{figure}
\centering
  \includegraphics[width=0.9\linewidth]{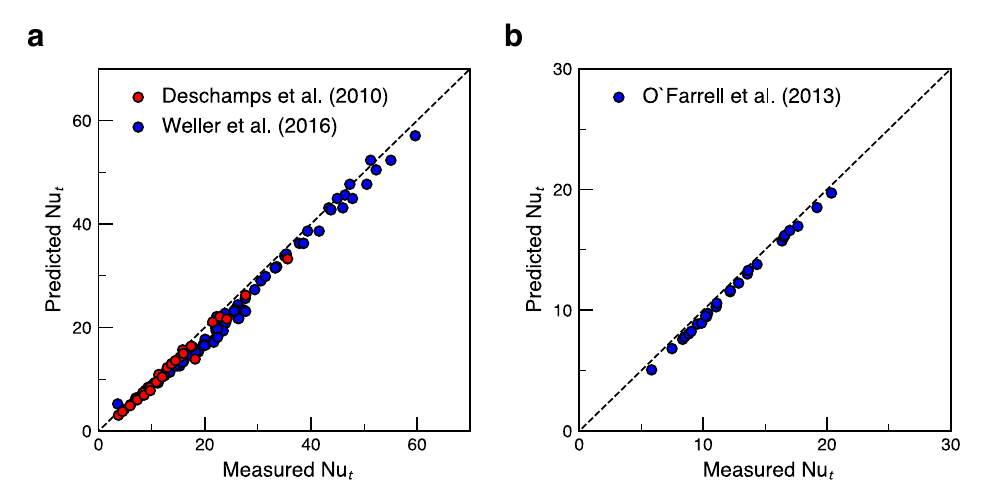}
  \caption{Comparison of previously published numerical simulations with the scaling for mixed heated convection in spherical geometry with (a) isoviscous rheology (eq.~\ref{eq:47}) and (b) depth-dependent rheology (eq.~\ref{eq:48}).}
  \label{fig:11}
\end{figure}

It has previously been demonstrated for the end-member heating cases that spherical geometry can be accounted for by incorporating a geometrical factor in the scaling laws for 2-D Cartesian geometry \cite<e.g.,>{vilella2017fully}. This is also true for convection in the mixed heating mode. A spherical shell domain can be characterized by $f$, the ratio of the inner radius to the outer radius. The greater surface area of the upper boundary with respect to the lower boundary means that, in order for energy to be conserved, the upper boundary must experience a lower heat flow per unit area than the lower boundary (at least in the case of no internal heating). In general, we must modify the heat conservation equation (eq.~\ref{eq:15}) as follows:
\begin{linenomath*}
\begin{equation}
    \label{eq:46}
    Nu_t = H^*\frac{1-f^3}{3\left(1-f \right)}+Nu_b f^2.
\end{equation}
\end{linenomath*}
As a result, the final scaling becomes
\begin{linenomath*}
\begin{equation}
    \label{eq:47}
    \left(\Delta T_t^{\mbox{\scriptsize HF}}\right)^{4/3} =f^2\left(\frac{1+\sigma}{b} - \Delta T_t^{\mbox{\scriptsize HF}}\right)^{4/3}+ \frac{H^*}{c}\frac{1-f^3}{3\left(1-f \right)}\left( b\frac{Ra}{Ra_{cr}} \right)^{1/3},
\end{equation}
\end{linenomath*}
where we may still use eqs.~\ref{eq:16}--\ref{eq:18} to solve for $Nu_t$ after obtaining $\Delta T_t^{\mbox{\scriptsize HF}}$. We use this scaling to predict $Nu_t$ in the numerical experiments of \citeA{deschamps2010temperature} and \citeA{weller2016scaling} for isoviscous convection in spherical geometry. While \citeA{deschamps2010temperature} normalize lengths using the thickness of the spherical shell, which is consistent with how our scaling is defined, \citeA{weller2016scaling} normalize lengths using the total radius of the outer boundary. Thus, before using our scaling to predict $Nu_t$, we first modify the values of $Ra$ and $H^*$ reported by \citeA{weller2016scaling} to account for this. Fig.~\ref{fig:11}a compares our scaling predictions with the measurements of \citeA{deschamps2010temperature} and \citeA{weller2016scaling}; the scaling is remarkably effective, considering that we have assumed the same $Ra_{cr}$, $b$, $c$, and $\sigma$ parameterizations derived for the 2-D Cartesian case.

We now turn to the case of a fluid with depth-dependent viscosity in a spherical shell domain, for which \citeA{o2013comparison} have performed numerical experiments. The viscosity structure used in their simulations consists of continuously increasing viscosity in the lower portion of the spherical shell, with a maximum nondimensional viscosity of 30 at the base. In order to make use of the scaling we have derived for layered viscosity in section 3.3, we will assume that the entire bottom TBL may be characterized by a viscosity of 30, which is reasonable in the limit of large $Ra$, for which TBLs are thin. We again make use of eq.~\ref{eq:46} to account for spherical geometry, to arrive at:
\begin{linenomath*}
\begin{equation}
    \label{eq:48}
    \left(\Delta T_t^{\mbox{\scriptsize HF}}\right)^{4/3} =f^2\left(\frac{1+\sigma}{b} - \Delta T_t^{\mbox{\scriptsize HF}}\right)^{4/3}\eta_{\mbox{\scriptsize max}}^{-1/3}+ \frac{H^*}{c}\frac{1-f^3}{3\left(1-f \right)}\left( b\frac{Ra}{Ra_{cr}} \right)^{1/3},
\end{equation}
\end{linenomath*}
where $\eta_{\mbox{\scriptsize max}} = 30$. Here, too, we use the same numerical constants determined for the 2D planar case, and the resulting predictions are successful (Fig.~\ref{fig:11}b).

\section{Discussion}
\subsection{Implications for global geodynamics and thermal evolution modeling}
Previous studies of convection in the mixed heating mode \cite{sotin1999three,moore2008heat,vilella2018temperature} suggested that interactions between the top and bottom boundary layer may invalidate the boundary layer stability criterion and thus its use for deriving scaling laws. We have shown that, as long as TBL interactions are appropriately accounted for (in our case, by describing the so-called temperature overshoot $\sigma$ of the TBLs), boundary layer stability analysis successfully describes mixed heated convection. This has allowed us to develop scaling laws based on the underlying physics, which lends confidence to the extension of such scaling laws to broader parameter spaces and to real-Earth complexities.

The question of whether heat flux and TBL properties are globally or locally determined has long remained nebulous \cite<e.g.,>{stevenson1983magnetism}. Thus, a key finding of our scaling analysis is that the surface heat flux is expected to depend only on the structure of the top TBL, and the basal heat flux only on the structure of the bottom TBL, not on the entire system. This agrees with what \citeA{howard1966convection} originally proposed, but how depth dependence of material properties affects the behavior and observable features of mantle convection is a question that has been around for a long time. For example, how depth dependence of viscosity influences the planform of convection has been unclear \cite{bunge1996effect,tackley1996ability}. While planform is somewhat of a secondary convective property, we have shown that how heat is transported at the surface depends only on the local structure of the TBL. Additionally, in order to reproduce Earth's measured heat flux with a simple scaling argument, very high viscosity is needed (e.g., $10^{22}$ Pa s), and it has often been thought that this may represent the lower mantle viscosity \cite<e.g.,>{bercovici2000relation,bercovici20157}. Under this scenario, the surface heat flux is dependent on the global distribution of material properties. This may appear reasonable, as the manner in which subducted material descends is likely regulated by lower mantle viscosity. Our scaling for depth-dependent viscosity suggests, however, that this high viscosity represents an effective lithospheric viscosity, as the surface heat flux is governed by properties of the upper thermal boundary layer (i.e., the lithosphere).

The fact that the boundary layer stability criterion is valid for mixed heating, and thus the surface heat flux is simply governed by the top TBL, means that thermal evolution modeling may proceed much as it has long been conducted. For example, modeling Earth’s thermal evolution backwards in time using our scaling laws would proceed as follows. First, one would use the dimensional version of eq.~\ref{eq:47} to solve for $H$, using estimates of the present-day thermal structure of the lithosphere as well as the Earth's $Ra$. Because secular cooling can be considered a contribution to internal heat generation for steady state solutions \cite<e.g.,>{korenaga2017pitfalls}, it may be solved for from $H$ by assuming the amount of radiogenic heat produced in the mantle. At each subsequent timestep, one would solve for the surface and core heat fluxes using equations similar to eq.~\ref{eq:49} (below) using the updated mantle temperature. Secular cooling is then simply found by balancing the surface heat flux with the core heat flux, radiogenic heat production, and secular cooling. Apart from numerically solving for $H$ at the initial timestep using some form of eq.~\ref{eq:48}, this approach is identical to how thermal evolution is traditionally modeled. Further, the temperature overshoot $\sigma$ only need be considered at the initial timestep in eq.~\ref{eq:47}. Since our scaling of $\sigma$ only depends on $Ra$, its incorporation is straightforward. It may seem like the use of $\sigma$ and eq.~\ref{eq:47} may not be so important, since the thermal evolution modeling proceeds as usual after the first time step; however, our scaling analysis shows that these components ensure modeling is conducted in a physically consistent manner. It is reassuring that traditional thermal evolution modeling is largely well-founded, as previous scaling analyses questioned the boundary layer stability criterion, the foundational assumption of such modeling.

\subsection{Application to lithospheric strength}
When applying our scaling theory to Earth, it is not immediately obvious that marginal stability applies to the entirety of the lithosphere. The so-called small-scale convection affects only the base of the lithosphere \cite{davaille1994onset,korenaga2003physics}, and this process resembles the stagnant lid mode of convection, where marginal stability only applies to a thin sublayer of the lithosphere. However, some weakening mechanism evidently allows for subduction of the lithosphere \cite{bercovici20157,korenaga2020plate}, and it is the marginal stability of the entire lithosphere that allows for this subduction and for the continuous operation of plate tectonics. Additionally, the lithosphere does not deform purely viscously; to incorporate the effect of plastic deformation into scaling laws for a viscous fluid, viscosity can be treated as an effective parameter \cite<e.g.,>{moresi1998mantle}.

With this in mind, our scaling analysis implies that the surface heat flux of Earth's mantle is simply governed by the marginal stability of lithosphere. Since we can reasonably estimate the heat flux coming out of the mantle, we may in theory infer lithospheric properties. In what follows, we attempt to estimate the effective viscosity of Earth's lithosphere.

By applying the dimensional versions of eqs.~\ref{eq:16a} and \ref{eq:28} to Earth's mantle, we arrive at:
\begin{linenomath*}
\begin{equation}
    \label{eq:49}
    \frac{Q_M}{\left(k \Delta T /D \right)} = 4\pi R_E^2 c \left(\frac{\Delta T_l}{\Delta T} \right)^{4/3} \left(b\frac{Ra}{Ra_{cr}} \right)^{1/3}\Delta \eta_l^{-1/3},
\end{equation}
\end{linenomath*}
where $R_E$ is the radius of Earth, $\Delta T_{l}$ is the temperature contrast across the lithosphere, $Ra$ is defined as in eq.~\ref{eq:4}, and $\Delta \eta_l = \eta_l / \eta_0$ is the viscosity contrast between the lithosphere and the convecting mantle. Actual viscosity varies greatly in the lithosphere, given its temperature dependence. Thus, the lithospheric viscosity $\eta_l$ is an effective viscosity that represents lithospheric stiffness with a single value.

Because we have reasonable estimates of $Q_M$ and $\Delta T_l$ (Table~\ref{table:5}), we can solve for $\Delta \eta_l$ in eq.~\ref{eq:49} by assuming some reference mantle viscosity $\eta_0$ to compute the Rayleigh number of the mantle. We test a range of values for $\eta_0$, as this parameter involves a high degree of uncertainty \cite<e.g.,>{forte2015constraints}.

\begin{figure}
\centering
  \includegraphics[width=0.9\linewidth]{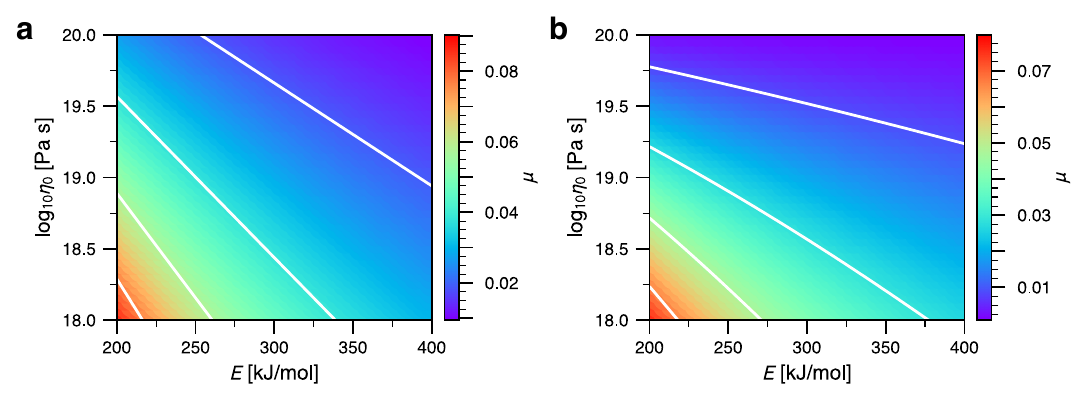}
  \caption{Application of our scaling assumptions to solve for the friction coefficient of Earth's lithosphere as a function of activation energy and mantle reference viscosity. In (A), the effect of dehydration stiffening is not considered, and in (B) this effect is considered. The values of $\mu$ shown in (B) are taken as minimum possible values, as we consider the extreme case of $\Delta \eta_D = 10^3$ and $h^*/h_{\mathrm{ref}}^* = 10$ for use in eq.~\ref{eq:51}. Solid white contour lines demarcate intervals of 0.02.}
  \label{fig:12}
\end{figure}

The scaling analysis of \citeA{korenaga2010scaling} suggests the following relationship between lithospheric viscosity contrast, lithospheric friction coefficient, and the Frank-Kamenetskii parameter:
\begin{linenomath*}
\begin{equation}
    \label{eq:50}
    \Delta \eta_l \left(\gamma, \theta \right) = \mathrm{exp}\left[A(\gamma)\theta \right],
\end{equation}
\end{linenomath*}
where $A(\gamma) = 0.327\gamma^{0.647}$, $\gamma = \mu/(\alpha \Delta T)$, and $\mu$ is the effective friction coefficient. If we assume some activation energy $E$ for the mantle, we may use eq.~\ref{eq:6b} to compute $\theta$ for the mantle, and in turn solve for $\mu$. We test a range of $E$, which is also not well constrained \cite{jain2020synergy}. Thus, we estimate $\mu$ as a function of both $\eta_0$ and $E$. The parameters assumed in this calculation are listed in Table~\ref{table:5}. In all cases, $\mu$ is small (less than 0.1; Fig.~\ref{fig:12}a), which is unsurprising given that the lithosphere must be weak enough to subduct. Both low $\eta_0$ and low $E$ contribute to a large $\mu$.
\begin{table}
\caption{Parameters used in the application of scaling assumptions to Earth's lithosphere}
\centering
\begin{tabular}{c c c}
\hline
Parameter & Unit & Value  \\
\hline
 $\alpha$ & K$^{-1}$ & $2\times 10^{-5}$\\
 $\rho_0$ & kg~m$^{-3}$ & $4500$\\
 $g$ & m~s$^{-2}$ & 9.8\\
 $\Delta T$$^\mathrm{a}$ & K & 1850 \\
 $D$ & m & $2.9\times 10^6$ \\
 $\kappa$ & m$^2$~s$^{-1}$ & $10^{-6}$ \\
 $\eta_0$ & Pa s & $10^{18}$ to $10^{20}$ \\
 $E$$^\mathrm{b}$ & kJ~mol$^{-1}$ & 200 to 400 \\
 $R$ & J~mol$^{-1}$~K$^{-1}$ & 8.3145 \\
 $T_S$ & K & $273$ \\
 $Q_M$$^\mathrm{c}$ & TW & $36$ \\
 $k$ & W~m$^{-1}$~K$^{-1}$ & $3$ \\
 $R_E$ & m & $6.37\times 10^{6}$ \\
 $\Delta T_l$$^\mathrm{d}$ & K & $1350$ \\
 $b$ & & $0.95$ \\
 $c$ & & $2.5$ \\
 $Ra_{cr}$ &  & $500$ \\
\hline
\end{tabular}

\footnotesize{$^\mathrm{a}$The sum of $\Delta T_l$ and the temperature jump across the lower mantle boundary layer, roughly 500 K \cite{deschamps2004towards}. $^\mathrm{b}$\citeA{hirth2003rheology,jain2019global}. $^\mathrm{c}$\citeA{jaupart2007heat}. $^\mathrm{d}$\citeA{herzberg2007temperatures}.}\\
\label{table:5}
\end{table}

We may also include the effect of dehydration stiffening that occurs as a result of mantle melting. This is formulated as \cite{korenaga2010scaling}:
\begin{linenomath*}
\begin{equation}
    \label{eq:51}
    \Delta \eta_l = \Delta \eta_{l,\mathrm{ref}}\mathrm{exp}\left[\mathrm{ln}\left(\Delta \eta_D\right) \mathrm{min}\left(1,\frac{h^*}{\chi h_{\mathrm{ref}}^*} \right) \right],
\end{equation}
\end{linenomath*}
where $\Delta \eta_{l,\mathrm{ref}}$ is the lithospheric viscosity contrast without considering dehydration stiffening (referred to as $\Delta \eta_l$ above), $\Delta \eta_D$ is the viscosity contrast due to dehydration, $\chi = 6$, and $h^*/h_{\mathrm{ref}}^*$ is the normalized thickness of the dehydrated layer. While $\Delta \eta_D$ and $h^*/h_{\mathrm{ref}}^*$ are relatively uncertain, we can investigate an extreme case to estimate the maximum effect on $\mu$. We choose $\Delta \eta_D = 10^3$ and $h^*/h_{\mathrm{ref}}^* = 10$ for this extreme case, and find that $\mu$ decreases slightly and is less than 0.08 (Fig.~\ref{fig:12}b).

\section{Conclusions}
We have derived scaling laws for convection in the mixed heating mode starting from the physics of such convection. These scaling laws succeed remarkably in predicting major convection diagnostics of numerical simulations, even when extended to depth-dependent viscosity, temperature-dependent viscosity, and spherical geometry. At the heart of our scaling analysis is the boundary layer stability criterion, the applicability of which has been questioned for mixed heated convection. The success of this criterion has important and encouraging implications. First, the heat flux at the surface and basal boundaries are determined locally by the thermal boundary layer structure and not globally. And second, the classical method of thermal evolution modeling is appropriate for determining the thermal history of terrestrial planets.

\appendix
\section{Parameterization of TBL temperature overshoot}
In section 3.2, we established that upwellings and downwellings may perturb the thermal structure of the opposite TBL, leading to an overshoot $\sigma$ equal to $\Delta T_t^{\mathrm{CR}}+\Delta T_b^{\mathrm{CR}} - 1$. Consider a downwelling parcel of fluid; its effect on the thermal structure of the opposite TBL depends on its temperature when it reaches the bottom TBL. The cold upper TBL has an average temperature of roughly $\Delta T_t / 2$, where $\Delta T_t$ is approximately the interior temperature, and we may assume that the downwelling is also characterized by this temperature when it initially detaches and starts to descend (call this initial temperature $T_i$). As it descends, its temperature increases by thermal diffusion: $\delta T/\delta t \propto \Delta T/\Delta x^2 + \Delta T/\Delta y^2$. Here, $\delta T$ is the temperature change of the parcel as it descends (such that the final parcel temperature $T_f$ when it reaches the bottom TBL is $T_i+\delta T$), $\delta t$ is the time it takes to descend, $\Delta T$ is the difference in temperature between the parcel and the ambient convecting interior, and $\Delta x$ and $\Delta y$ are the size of the parcel in the $x$ and $y$ dimensions, respectively. The term $\Delta T/\Delta y^2$ can be neglected because the parcel is a thin, long, and vertically-oriented structure (see for example Fig.~\ref{fig:2}b), such that $\Delta y$ is large. The downwelling time, $\delta t$, will depend on vertical velocity $w$ and the distance travelled by the parcel before reaching the bottom TBL. Because the TBLs are thin (in the limit of high $Ra$) this distance is approximately $1$, the total height of the system. Thus, $\delta t \sim 1/w$. We can approximate $\Delta x$, the thickness of the downwelling parcel, by considering that the downwelling originates from the top TBL. The size of the downwelling will be proportional to the thickness of the top TBL: $\Delta x \propto \delta_t$. Next, recalling that the initial parcel temperature is roughly $\Delta T_t/2$, the difference between the parcel temperature and the interior temperature (approximately $\Delta T_t$) will be proportional to $\Delta T_t$ itself. We can reformulate $\Delta T_t$ as $Nu_t \delta_t$ using eq.~\ref{eq:16a}, so that we finally arrive at $\delta T \propto Nu_t/(\delta_t w)$. Thus, the parcel temperature when it arrives at the bottom TBL is $T_f = T_i+\delta T \propto \Delta T_t/2 + Nu_t/(\delta_t w)$. The temperature anomaly caused by the downwelling is given by the difference between $T_f$ and the temperature of the bottom TBL near its inner boundary. At the upper boundary of the bottom TBL, the unperturbed temperature will be roughly equal to the internal temperature (approximated by $\Delta T_t$). Thus, the temperature anomaly from the downwelling is proportional to $-\Delta T_t/2 - Nu_t/(\delta_t w)$. This quantity is negative because we assume that the vertical velocity is large enough so that the parcel is still colder than its surroundings when it reaches the bottom TBL. If we further assume that $\Delta T_t$ is roughly $1/2$ (this is true for cases with low internal heating ratio), then we can simplify this quantity to $C - Nu_t/(\delta_t w)$, where $C$ is some constant. To determine the overshoot in the horizontally averaged temperature profile, we need to multiply this quantity by $\delta_t$. This is because we need to integrate over the size of the parcel to determine the perturbation of the averaged profile. We can justify this factor of $\delta_t$ as follows.

Consider the thermal structure at a single timestep (such as in Fig.~\ref{fig:2}b) and at a single height $y=y^*$ near the inner boundary of the bottom TBL where the temperature overshoot is prominent. The horizontally averaged temperature at $y=y^*$ is given by
\begin{linenomath*}
\begin{equation}
    \label{eq:A1}
    \overline{T}(y=y^*) = \frac{1}{L}\int^L_0 T (x,y=y^*) dx,
\end{equation}
\end{linenomath*}
where $L$ is the nondimensional horizontal length of the domain (in the case of our numerical simulations, $L=4$). If we assume that some length $X$ of $T(x,y=y^*)$ is characterized by the anomalous temperature $T_f$ due to an arriving downwelling, and the rest of the material at $y=y^*$ is characterized by the ambient temperature (approximate this as $\Delta T_t$ since $y^*$ is the near the convecting interior), then we have
\begin{linenomath*}
\begin{equation}
    \label{eq:A2}
    \overline{T}(y=y^*) = \frac{1}{L}\left[ \int^X_0 T_f dx + \int_X^L \Delta T_t dx \right] = \frac{1}{L}\left(T_f X + \Delta T_t\left(L-X \right) \right).
\end{equation}
\end{linenomath*}
It is reasonable to assume that the length $X$ characterized by the anomalous temperature should be proportional to the size of the downwelling, which can be approximated by $\delta_t$. Thus,
\begin{linenomath*}
\begin{equation}
    \label{eq:A3}
    \overline{T}(y=y^*) = \Delta T_t + \frac{1}{L}\delta_t\left(T_f-\Delta T_t\right).
\end{equation}
\end{linenomath*}
Because the ambient temperature at $y=y^*$ is $\Delta T_t$, the deviation from this temperature, $\frac{1}{L}\delta_t\left(T_f-\Delta T_t\right)$, is the overshoot itself. It was determined above that $T_f-\Delta T_t \propto C-Nu_t/(\delta_t w)$, So to obtain the overshoot in the horizontally averaged temperature profile, this quantity must be multiplied by a factor proportional to $\delta_t$.

\begin{figure}
\centering
  \includegraphics[width=0.6\linewidth]{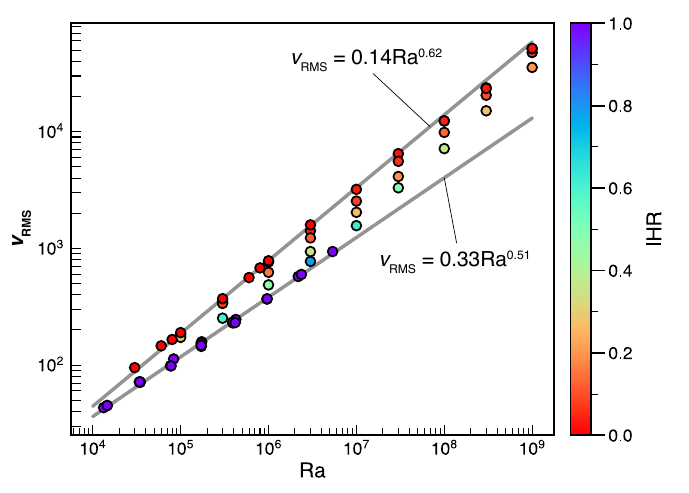}
  \caption{Measured root-mean-square (RMS) velocity as a function of $Ra$ in the numerical simulations. Circles are colored by internal heating ratio (IHR), defined as $H^*/Nu_t$. In addition to the runs listed in Table~\ref{table:1}, we have included a number of purely internally heated runs, in which case the input Rayleigh number must be rescaled using the a posteriori maximum temperature $T_{\mathrm{max}}$ of the system (i.e., we plot $RaT_{\mathrm{max}}$ on the x-axis).}
  \label{fig:A1}
\end{figure}

As a result, the overshoot due to the downwelling parcel is proportional to $C\delta_t - Nu_t/w$. To convert this quantity to a function of $Ra$ and/or $H^*$, we consider the limit of Rayleigh-B\'enard convection, which has well-defined scalings for $\delta_t$ and $Nu_t$, which are proportional to $Ra^{-1/3}$ and $Ra^{1/3}$, respectively. Lastly, we assume $w \propto Ra^{0.55}$. It is well known that convective velocities depend strongly on $Ra$, and the exponent 0.55 is roughly midway between the exponents measured for purely internally heated runs and purely basally heated runs (Fig.~\ref{fig:A1}). Thus, these considerations suggest that the overshoot caused by downwellings is proportional to $C'Ra^{-1/3} - Ra^{-0.22}$. We can do a similar analysis for the effect of upwellings on the temperature structure of the top TBL, and find an overshoot proportional to $-C'Ra^{-1/3} + Ra^{-0.22}$. Collectively, the scaling for the overshoot is given by $\sigma = c_1 Ra^{-1/3}+c_2 Ra^{-0.22}$, where $c_1$ and $c_2$ are unknown constants. Upon comparison with numerical experiments, we find that $c_1 = -10.39$ and $c_2 = 4.01$ are the best-fit constants (Fig.~\ref{fig:3}), resulting in eq.~\ref{eq:20}.

%



%
%

\section*{Open Research Section}
This work is theoretical in nature and can be reproduced from the methods described in the text. All numerical data are presented in Tables \ref{table:1}, \ref{table:3}, and \ref{table:4} and can be accessed directly at doi.org/10.17632/c95ysmspfm.1 \cite{Ferrick2023}.

\acknowledgments
This work is supported by the U.S. National Science Foundation grant EAR-1753916 (J.K.). The authors thank two anonymous reviewers for insightful and constructive comments.

%
%

\bibliography{references.bib}

\begin{thebibliography}{}

\bibitem [\protect \citeauthoryear {%
Bercovici%
, Ricard%
\BCBL {}\ \BBA {} Richards%
}{%
Bercovici%
\ \protect \BOthers {.}}{%
{\protect \APACyear {2000}}%
}]{%
bercovici2000relation}
\APACinsertmetastar {%
bercovici2000relation}%
\begin{APACrefauthors}%
Bercovici, D.%
, Ricard, Y.%
\BCBL {}\ \BBA {} Richards, M\BPBI A.%
\end{APACrefauthors}%
\unskip\
\newblock
\APACrefYearMonthDay{2000}{}{}.
\newblock
{\BBOQ}\APACrefatitle {The relation between mantle dynamics and plate
  tectonics: A primer} {The relation between mantle dynamics and plate
  tectonics: A primer}.{\BBCQ}
\newblock
\BIn{} M\BPBI A.~Richards, R\BPBI G.~Gordon\BCBL {}\ \BBA {} R\BPBI D.~van~der
  Hilst\ (\BEDS), \APACrefbtitle {The {H}istory and {D}ynamics of {G}lobal
  {P}late {M}otions} {The {H}istory and {D}ynamics of {G}lobal {P}late
  {M}otions}\ (\BPGS\ 5--46).
\newblock
\APACaddressPublisher{Washington, D.C.}{AGU}.
\PrintBackRefs{\CurrentBib}

\bibitem [\protect \citeauthoryear {%
Bercovici%
, Schubert%
\BCBL {}\ \BBA {} Glatzmaier%
}{%
Bercovici%
\ \protect \BOthers {.}}{%
{\protect \APACyear {1992}}%
}]{%
bercovici1992three}
\APACinsertmetastar {%
bercovici1992three}%
\begin{APACrefauthors}%
Bercovici, D.%
, Schubert, G.%
\BCBL {}\ \BBA {} Glatzmaier, G\BPBI A.%
\end{APACrefauthors}%
\unskip\
\newblock
\APACrefYearMonthDay{1992}{}{}.
\newblock
{\BBOQ}\APACrefatitle {Three-dimensional convection of an
  infinite-{P}randtl-number compressible fluid in a basally heated spherical
  shell} {Three-dimensional convection of an infinite-{P}randtl-number
  compressible fluid in a basally heated spherical shell}.{\BBCQ}
\newblock
\APACjournalVolNumPages{J. Fluid Mech.}{239}{}{683--719}.
\PrintBackRefs{\CurrentBib}

\bibitem [\protect \citeauthoryear {%
Bercovici%
, Tackley%
\BCBL {}\ \BBA {} Ricard%
}{%
Bercovici%
\ \protect \BOthers {.}}{%
{\protect \APACyear {2015}}%
}]{%
bercovici20157}
\APACinsertmetastar {%
bercovici20157}%
\begin{APACrefauthors}%
Bercovici, D.%
, Tackley, P.%
\BCBL {}\ \BBA {} Ricard, Y.%
\end{APACrefauthors}%
\unskip\
\newblock
\APACrefYearMonthDay{2015}{}{}.
\newblock
{\BBOQ}\APACrefatitle {The generation of plate tectonics from mantle dynamics}
  {The generation of plate tectonics from mantle dynamics}.{\BBCQ}
\newblock
\BIn{} G.~Schubert\ (\BED), \APACrefbtitle {Treatise on {G}eophysics, 2nd ed.}
  {Treatise on {G}eophysics, 2nd ed.}\ (\BVOL~7, \BPGS\ 271--318).
\newblock
\APACaddressPublisher{Oxford}{Elsevier}.
\PrintBackRefs{\CurrentBib}

\bibitem [\protect \citeauthoryear {%
Bunge%
, Richards%
\BCBL {}\ \BBA {} Baumgardner%
}{%
Bunge%
\ \protect \BOthers {.}}{%
{\protect \APACyear {1996}}%
}]{%
bunge1996effect}
\APACinsertmetastar {%
bunge1996effect}%
\begin{APACrefauthors}%
Bunge, H\BHBI P.%
, Richards, M\BPBI A.%
\BCBL {}\ \BBA {} Baumgardner, J\BPBI R.%
\end{APACrefauthors}%
\unskip\
\newblock
\APACrefYearMonthDay{1996}{}{}.
\newblock
{\BBOQ}\APACrefatitle {Effect of depth-dependent viscosity on the planform of
  mantle convection} {Effect of depth-dependent viscosity on the planform of
  mantle convection}.{\BBCQ}
\newblock
\APACjournalVolNumPages{Nature}{379}{6564}{436--438}.
\PrintBackRefs{\CurrentBib}

\bibitem [\protect \citeauthoryear {%
Christensen%
}{%
Christensen%
}{%
{\protect \APACyear {1984}}%
}]{%
christensen1984convection}
\APACinsertmetastar {%
christensen1984convection}%
\begin{APACrefauthors}%
Christensen, U.%
\end{APACrefauthors}%
\unskip\
\newblock
\APACrefYearMonthDay{1984}{}{}.
\newblock
{\BBOQ}\APACrefatitle {Convection with pressure- and temperature-dependent
  non-{N}ewtonian rheology} {Convection with pressure- and
  temperature-dependent non-{N}ewtonian rheology}.{\BBCQ}
\newblock
\APACjournalVolNumPages{Geophys. J. Int.}{77}{2}{343--384}.
\PrintBackRefs{\CurrentBib}

\bibitem [\protect \citeauthoryear {%
Christensen%
}{%
Christensen%
}{%
{\protect \APACyear {1985}}%
}]{%
christensen1985thermal}
\APACinsertmetastar {%
christensen1985thermal}%
\begin{APACrefauthors}%
Christensen, U\BPBI R.%
\end{APACrefauthors}%
\unskip\
\newblock
\APACrefYearMonthDay{1985}{}{}.
\newblock
{\BBOQ}\APACrefatitle {Thermal evolution models for the {E}arth} {Thermal
  evolution models for the {E}arth}.{\BBCQ}
\newblock
\APACjournalVolNumPages{J. Geophys. Res.}{90}{B4}{2995--3007}.
\PrintBackRefs{\CurrentBib}

\bibitem [\protect \citeauthoryear {%
Davaille%
\ \BBA {} Jaupart%
}{%
Davaille%
\ \BBA {} Jaupart%
}{%
{\protect \APACyear {1993}}%
}]{%
davaille1993transient}
\APACinsertmetastar {%
davaille1993transient}%
\begin{APACrefauthors}%
Davaille, A.%
\BCBT {}\ \BBA {} Jaupart, C.%
\end{APACrefauthors}%
\unskip\
\newblock
\APACrefYearMonthDay{1993}{}{}.
\newblock
{\BBOQ}\APACrefatitle {Transient high-{R}ayleigh-number thermal convection with
  large viscosity variations} {Transient high-{R}ayleigh-number thermal
  convection with large viscosity variations}.{\BBCQ}
\newblock
\APACjournalVolNumPages{J. Fluid Mech.}{253}{}{141--166}.
\PrintBackRefs{\CurrentBib}

\bibitem [\protect \citeauthoryear {%
Davaille%
\ \BBA {} Jaupart%
}{%
Davaille%
\ \BBA {} Jaupart%
}{%
{\protect \APACyear {1994}}%
}]{%
davaille1994onset}
\APACinsertmetastar {%
davaille1994onset}%
\begin{APACrefauthors}%
Davaille, A.%
\BCBT {}\ \BBA {} Jaupart, C.%
\end{APACrefauthors}%
\unskip\
\newblock
\APACrefYearMonthDay{1994}{}{}.
\newblock
{\BBOQ}\APACrefatitle {Onset of thermal convection in fluids with
  temperature-dependent viscosity: Application to the oceanic mantle} {Onset of
  thermal convection in fluids with temperature-dependent viscosity:
  Application to the oceanic mantle}.{\BBCQ}
\newblock
\APACjournalVolNumPages{J. Geophys. Res.}{99}{B10}{19853--19866}.
\PrintBackRefs{\CurrentBib}

\bibitem [\protect \citeauthoryear {%
Deschamps%
, Tackley%
\BCBL {}\ \BBA {} Nakagawa%
}{%
Deschamps%
\ \protect \BOthers {.}}{%
{\protect \APACyear {2010}}%
}]{%
deschamps2010temperature}
\APACinsertmetastar {%
deschamps2010temperature}%
\begin{APACrefauthors}%
Deschamps, F.%
, Tackley, P\BPBI J.%
\BCBL {}\ \BBA {} Nakagawa, T.%
\end{APACrefauthors}%
\unskip\
\newblock
\APACrefYearMonthDay{2010}{}{}.
\newblock
{\BBOQ}\APACrefatitle {Temperature and heat flux scalings for isoviscous
  thermal convection in spherical geometry} {Temperature and heat flux scalings
  for isoviscous thermal convection in spherical geometry}.{\BBCQ}
\newblock
\APACjournalVolNumPages{Geophys. J. Int.}{182}{1}{137--154}.
\PrintBackRefs{\CurrentBib}

\bibitem [\protect \citeauthoryear {%
Deschamps%
\ \BBA {} Trampert%
}{%
Deschamps%
\ \BBA {} Trampert%
}{%
{\protect \APACyear {2004}}%
}]{%
deschamps2004towards}
\APACinsertmetastar {%
deschamps2004towards}%
\begin{APACrefauthors}%
Deschamps, F.%
\BCBT {}\ \BBA {} Trampert, J.%
\end{APACrefauthors}%
\unskip\
\newblock
\APACrefYearMonthDay{2004}{}{}.
\newblock
{\BBOQ}\APACrefatitle {Towards a lower mantle reference temperature and
  composition} {Towards a lower mantle reference temperature and
  composition}.{\BBCQ}
\newblock
\APACjournalVolNumPages{Earth Planet. Sci. Lett.}{222}{1}{161--175}.
\PrintBackRefs{\CurrentBib}

\bibitem [\protect \citeauthoryear {%
Ferrick%
}{%
Ferrick%
}{%
{\protect \APACyear {2023}}%
}]{%
Ferrick2023}
\APACinsertmetastar {%
Ferrick2023}%
\begin{APACrefauthors}%
Ferrick, A.%
\end{APACrefauthors}%
\unskip\
\newblock
\APACrefYearMonthDay{2023}{}{}.
\newblock
\APACrefbtitle {Data for: Generalizing scaling laws for mantle convection with
  mixed heating.} {Data for: Generalizing scaling laws for mantle convection
  with mixed heating.}
\newblock
\APACaddressPublisher{}{Mendeley Data}.
\newblock
\APACrefnote{V1}
\newblock
\begin{APACrefDOI} \doi{10.17632/c95ysmspfm.1} \end{APACrefDOI}
\PrintBackRefs{\CurrentBib}

\bibitem [\protect \citeauthoryear {%
Forte%
, Simmons%
\BCBL {}\ \BBA {} Grand%
}{%
Forte%
\ \protect \BOthers {.}}{%
{\protect \APACyear {2015}}%
}]{%
forte2015constraints}
\APACinsertmetastar {%
forte2015constraints}%
\begin{APACrefauthors}%
Forte, A\BPBI M.%
, Simmons, N\BPBI A.%
\BCBL {}\ \BBA {} Grand, S\BPBI P.%
\end{APACrefauthors}%
\unskip\
\newblock
\APACrefYearMonthDay{2015}{}{}.
\newblock
{\BBOQ}\APACrefatitle {Constraints on 3-{D} seismic models from global
  geodynamic observables: Implications for the global mantle convective flow}
  {Constraints on 3-{D} seismic models from global geodynamic observables:
  Implications for the global mantle convective flow}.{\BBCQ}
\newblock
\BIn{} B.~Romanowicz\ \BBA {} A.~Dziewonski\ (\BEDS), \APACrefbtitle {Treatise
  on {G}eophysics, 2nd ed.} {Treatise on {G}eophysics, 2nd ed.}\ (\BVOL~1,
  \BPGS\ 853--907).
\newblock
\APACaddressPublisher{Oxford}{Elsevier}.
\PrintBackRefs{\CurrentBib}

\bibitem [\protect \citeauthoryear {%
Grasset%
\ \BBA {} Parmentier%
}{%
Grasset%
\ \BBA {} Parmentier%
}{%
{\protect \APACyear {1998}}%
}]{%
grasset1998thermal}
\APACinsertmetastar {%
grasset1998thermal}%
\begin{APACrefauthors}%
Grasset, O.%
\BCBT {}\ \BBA {} Parmentier, E.%
\end{APACrefauthors}%
\unskip\
\newblock
\APACrefYearMonthDay{1998}{}{}.
\newblock
{\BBOQ}\APACrefatitle {Thermal convection in a volumetrically heated, infinite
  {P}randtl number fluid with strongly temperature-dependent viscosity:
  Implications for planetary thermal evolution} {Thermal convection in a
  volumetrically heated, infinite {P}randtl number fluid with strongly
  temperature-dependent viscosity: Implications for planetary thermal
  evolution}.{\BBCQ}
\newblock
\APACjournalVolNumPages{J. Geophys. Res.}{103}{B8}{18171--18181}.
\PrintBackRefs{\CurrentBib}

\bibitem [\protect \citeauthoryear {%
Herzberg%
\ \protect \BOthers {.}}{%
Herzberg%
\ \protect \BOthers {.}}{%
{\protect \APACyear {2007}}%
}]{%
herzberg2007temperatures}
\APACinsertmetastar {%
herzberg2007temperatures}%
\begin{APACrefauthors}%
Herzberg, C.%
, Asimow, P\BPBI D.%
, Arndt, N.%
, Niu, Y.%
, Lesher, C.%
, Fitton, J.%
\BDBL {}Saunders, A.%
\end{APACrefauthors}%
\unskip\
\newblock
\APACrefYearMonthDay{2007}{}{}.
\newblock
{\BBOQ}\APACrefatitle {Temperatures in ambient mantle and plumes: Constraints
  from basalts, picrites, and komatiites} {Temperatures in ambient mantle and
  plumes: Constraints from basalts, picrites, and komatiites}.{\BBCQ}
\newblock
\APACjournalVolNumPages{Geochem. Geophys. Geosyst.}{8}{2}{Q02006}.
\PrintBackRefs{\CurrentBib}

\bibitem [\protect \citeauthoryear {%
Hirth%
\ \BBA {} Kohlstedt%
}{%
Hirth%
\ \BBA {} Kohlstedt%
}{%
{\protect \APACyear {2003}}%
}]{%
hirth2003rheology}
\APACinsertmetastar {%
hirth2003rheology}%
\begin{APACrefauthors}%
Hirth, G.%
\BCBT {}\ \BBA {} Kohlstedt, D.%
\end{APACrefauthors}%
\unskip\
\newblock
\APACrefYearMonthDay{2003}{}{}.
\newblock
{\BBOQ}\APACrefatitle {Rheology of the upper mantle and the mantle wedge: A
  view from the experimentalists} {Rheology of the upper mantle and the mantle
  wedge: A view from the experimentalists}.{\BBCQ}
\newblock
\BIn{} J.~Eiler\ (\BED), \APACrefbtitle {Inside the Subduction Factory} {Inside
  the subduction factory}\ (\BPGS\ 83--106).
\newblock
\APACaddressPublisher{Washington, D.C.}{AGU}.
\PrintBackRefs{\CurrentBib}

\bibitem [\protect \citeauthoryear {%
Howard%
}{%
Howard%
}{%
{\protect \APACyear {1966}}%
}]{%
howard1966convection}
\APACinsertmetastar {%
howard1966convection}%
\begin{APACrefauthors}%
Howard, L\BPBI N.%
\end{APACrefauthors}%
\unskip\
\newblock
\APACrefYearMonthDay{1966}{}{}.
\newblock
{\BBOQ}\APACrefatitle {Convection at high {R}ayleigh number} {Convection at
  high {R}ayleigh number}.{\BBCQ}
\newblock
\BIn{} H.~Gortler\ (\BED), \APACrefbtitle {Proceedings of the {E}leventh
  {I}nternational {C}ongress of {A}pplied {M}echanics} {Proceedings of the
  {E}leventh {I}nternational {C}ongress of {A}pplied {M}echanics}\ (\BPGS\
  1109--1115).
\newblock
\APACaddressPublisher{New York}{Springer}.
\PrintBackRefs{\CurrentBib}

\bibitem [\protect \citeauthoryear {%
Jain%
\ \BBA {} Korenaga%
}{%
Jain%
\ \BBA {} Korenaga%
}{%
{\protect \APACyear {2020}}%
}]{%
jain2020synergy}
\APACinsertmetastar {%
jain2020synergy}%
\begin{APACrefauthors}%
Jain, C.%
\BCBT {}\ \BBA {} Korenaga, J.%
\end{APACrefauthors}%
\unskip\
\newblock
\APACrefYearMonthDay{2020}{}{}.
\newblock
{\BBOQ}\APACrefatitle {Synergy of experimental rock mechanics, seismology, and
  geodynamics reveals still elusive upper mantle rheology} {Synergy of
  experimental rock mechanics, seismology, and geodynamics reveals still
  elusive upper mantle rheology}.{\BBCQ}
\newblock
\APACjournalVolNumPages{J. Geophys. Res. Solid Earth}{125}{11}{e2020JB019896}.
\PrintBackRefs{\CurrentBib}

\bibitem [\protect \citeauthoryear {%
Jain%
, Korenaga%
\BCBL {}\ \BBA {} Karato%
}{%
Jain%
\ \protect \BOthers {.}}{%
{\protect \APACyear {2019}}%
}]{%
jain2019global}
\APACinsertmetastar {%
jain2019global}%
\begin{APACrefauthors}%
Jain, C.%
, Korenaga, J.%
\BCBL {}\ \BBA {} Karato, S\BHBI i.%
\end{APACrefauthors}%
\unskip\
\newblock
\APACrefYearMonthDay{2019}{}{}.
\newblock
{\BBOQ}\APACrefatitle {Global analysis of experimental data on the rheology of
  olivine aggregates} {Global analysis of experimental data on the rheology of
  olivine aggregates}.{\BBCQ}
\newblock
\APACjournalVolNumPages{J. Geophys. Res. Solid Earth}{124}{1}{310--334}.
\PrintBackRefs{\CurrentBib}

\bibitem [\protect \citeauthoryear {%
Jarvis%
\ \BBA {} Mckenzie%
}{%
Jarvis%
\ \BBA {} Mckenzie%
}{%
{\protect \APACyear {1980}}%
}]{%
jarvis1980convection}
\APACinsertmetastar {%
jarvis1980convection}%
\begin{APACrefauthors}%
Jarvis, G\BPBI T.%
\BCBT {}\ \BBA {} Mckenzie, D\BPBI P.%
\end{APACrefauthors}%
\unskip\
\newblock
\APACrefYearMonthDay{1980}{}{}.
\newblock
{\BBOQ}\APACrefatitle {Convection in a compressible fluid with infinite
  {P}randtl number} {Convection in a compressible fluid with infinite {P}randtl
  number}.{\BBCQ}
\newblock
\APACjournalVolNumPages{J. Fluid Mech.}{96}{3}{515--583}.
\PrintBackRefs{\CurrentBib}

\bibitem [\protect \citeauthoryear {%
Jarvis%
\ \BBA {} Peltier%
}{%
Jarvis%
\ \BBA {} Peltier%
}{%
{\protect \APACyear {1982}}%
}]{%
jarvis1982mantle}
\APACinsertmetastar {%
jarvis1982mantle}%
\begin{APACrefauthors}%
Jarvis, G\BPBI T.%
\BCBT {}\ \BBA {} Peltier, W.%
\end{APACrefauthors}%
\unskip\
\newblock
\APACrefYearMonthDay{1982}{}{}.
\newblock
{\BBOQ}\APACrefatitle {Mantle convection as a boundary layer phenomenon}
  {Mantle convection as a boundary layer phenomenon}.{\BBCQ}
\newblock
\APACjournalVolNumPages{Geophys. J. Int.}{68}{2}{389--427}.
\PrintBackRefs{\CurrentBib}

\bibitem [\protect \citeauthoryear {%
Jaupart%
, Labrosse%
\BCBL {}\ \BBA {} Mareschal%
}{%
Jaupart%
\ \protect \BOthers {.}}{%
{\protect \APACyear {2007}}%
}]{%
jaupart2007heat}
\APACinsertmetastar {%
jaupart2007heat}%
\begin{APACrefauthors}%
Jaupart, C.%
, Labrosse, S.%
\BCBL {}\ \BBA {} Mareschal, J\BHBI C.%
\end{APACrefauthors}%
\unskip\
\newblock
\APACrefYearMonthDay{2007}{}{}.
\newblock
{\BBOQ}\APACrefatitle {Temperatures, heat and energy in the mantle of the
  {Earth}} {Temperatures, heat and energy in the mantle of the {Earth}}.{\BBCQ}
\newblock
\BIn{} G.~Schubert\ (\BED), \APACrefbtitle {Treatise on {G}eophysics} {Treatise
  on {G}eophysics}\ (\BVOL~7, \BPGS\ 253--303).
\newblock
\APACaddressPublisher{Amsterdam}{Elsevier}.
\PrintBackRefs{\CurrentBib}

\bibitem [\protect \citeauthoryear {%
Korenaga%
}{%
Korenaga%
}{%
{\protect \APACyear {2009}}%
}]{%
korenaga2009scaling}
\APACinsertmetastar {%
korenaga2009scaling}%
\begin{APACrefauthors}%
Korenaga, J.%
\end{APACrefauthors}%
\unskip\
\newblock
\APACrefYearMonthDay{2009}{}{}.
\newblock
{\BBOQ}\APACrefatitle {Scaling of stagnant-lid convection with {A}rrhenius
  rheology and the effects of mantle melting} {Scaling of stagnant-lid
  convection with {A}rrhenius rheology and the effects of mantle
  melting}.{\BBCQ}
\newblock
\APACjournalVolNumPages{Geophy. J. Int.}{179}{1}{154--170}.
\PrintBackRefs{\CurrentBib}

\bibitem [\protect \citeauthoryear {%
Korenaga%
}{%
Korenaga%
}{%
{\protect \APACyear {2010}}%
}]{%
korenaga2010scaling}
\APACinsertmetastar {%
korenaga2010scaling}%
\begin{APACrefauthors}%
Korenaga, J.%
\end{APACrefauthors}%
\unskip\
\newblock
\APACrefYearMonthDay{2010}{}{}.
\newblock
{\BBOQ}\APACrefatitle {Scaling of plate tectonic convection with pseudoplastic
  rheology} {Scaling of plate tectonic convection with pseudoplastic
  rheology}.{\BBCQ}
\newblock
\APACjournalVolNumPages{J. Geophys. Res.}{115}{}{B11405,}.
\PrintBackRefs{\CurrentBib}

\bibitem [\protect \citeauthoryear {%
Korenaga%
}{%
Korenaga%
}{%
{\protect \APACyear {2017}}%
}]{%
korenaga2017pitfalls}
\APACinsertmetastar {%
korenaga2017pitfalls}%
\begin{APACrefauthors}%
Korenaga, J.%
\end{APACrefauthors}%
\unskip\
\newblock
\APACrefYearMonthDay{2017}{}{}.
\newblock
{\BBOQ}\APACrefatitle {Pitfalls in modeling mantle convection with internal
  heat production} {Pitfalls in modeling mantle convection with internal heat
  production}.{\BBCQ}
\newblock
\APACjournalVolNumPages{J. Geophys. Res. Solid Earth}{122}{5}{4064--4085}.
\PrintBackRefs{\CurrentBib}

\bibitem [\protect \citeauthoryear {%
Korenaga%
}{%
Korenaga%
}{%
{\protect \APACyear {2020}}%
}]{%
korenaga2020plate}
\APACinsertmetastar {%
korenaga2020plate}%
\begin{APACrefauthors}%
Korenaga, J.%
\end{APACrefauthors}%
\unskip\
\newblock
\APACrefYearMonthDay{2020}{}{}.
\newblock
{\BBOQ}\APACrefatitle {Plate tectonics and surface environment: Role of the
  oceanic upper mantle} {Plate tectonics and surface environment: Role of the
  oceanic upper mantle}.{\BBCQ}
\newblock
\APACjournalVolNumPages{Earth Sci. Rev.}{205}{}{103185}.
\PrintBackRefs{\CurrentBib}

\bibitem [\protect \citeauthoryear {%
Korenaga%
\ \BBA {} Jordan%
}{%
Korenaga%
\ \BBA {} Jordan%
}{%
{\protect \APACyear {2003}}%
}]{%
korenaga2003physics}
\APACinsertmetastar {%
korenaga2003physics}%
\begin{APACrefauthors}%
Korenaga, J.%
\BCBT {}\ \BBA {} Jordan, T\BPBI H.%
\end{APACrefauthors}%
\unskip\
\newblock
\APACrefYearMonthDay{2003}{}{}.
\newblock
{\BBOQ}\APACrefatitle {Physics of multiscale convection in {E}arth's mantle:
  Onset of sublithospheric convection} {Physics of multiscale convection in
  {E}arth's mantle: Onset of sublithospheric convection}.{\BBCQ}
\newblock
\APACjournalVolNumPages{J. Geophys. Res.}{108}{B7}{2333}.
\PrintBackRefs{\CurrentBib}

\bibitem [\protect \citeauthoryear {%
Liu%
\ \BBA {} Zhong%
}{%
Liu%
\ \BBA {} Zhong%
}{%
{\protect \APACyear {2013}}%
}]{%
liu2013analyses}
\APACinsertmetastar {%
liu2013analyses}%
\begin{APACrefauthors}%
Liu, X.%
\BCBT {}\ \BBA {} Zhong, S.%
\end{APACrefauthors}%
\unskip\
\newblock
\APACrefYearMonthDay{2013}{}{}.
\newblock
{\BBOQ}\APACrefatitle {Analyses of marginal stability, heat transfer and
  boundary layer properties for thermal convection in a compressible fluid with
  infinite {P}randtl number} {Analyses of marginal stability, heat transfer and
  boundary layer properties for thermal convection in a compressible fluid with
  infinite {P}randtl number}.{\BBCQ}
\newblock
\APACjournalVolNumPages{Geophys. J. Int.}{194}{1}{125--144}.
\PrintBackRefs{\CurrentBib}

\bibitem [\protect \citeauthoryear {%
Moore%
}{%
Moore%
}{%
{\protect \APACyear {2008}}%
}]{%
moore2008heat}
\APACinsertmetastar {%
moore2008heat}%
\begin{APACrefauthors}%
Moore, W\BPBI B.%
\end{APACrefauthors}%
\unskip\
\newblock
\APACrefYearMonthDay{2008}{}{}.
\newblock
{\BBOQ}\APACrefatitle {Heat transport in a convecting layer heated from within
  and below} {Heat transport in a convecting layer heated from within and
  below}.{\BBCQ}
\newblock
\APACjournalVolNumPages{J. Geophys. Res.}{113}{}{B11407}.
\PrintBackRefs{\CurrentBib}

\bibitem [\protect \citeauthoryear {%
Moresi%
\ \BBA {} Solomatov%
}{%
Moresi%
\ \BBA {} Solomatov%
}{%
{\protect \APACyear {1998}}%
}]{%
moresi1998mantle}
\APACinsertmetastar {%
moresi1998mantle}%
\begin{APACrefauthors}%
Moresi, L.%
\BCBT {}\ \BBA {} Solomatov, V.%
\end{APACrefauthors}%
\unskip\
\newblock
\APACrefYearMonthDay{1998}{}{}.
\newblock
{\BBOQ}\APACrefatitle {Mantle convection with a brittle lithosphere: thoughts
  on the global tectonic styles of the {E}arth and {V}enus} {Mantle convection
  with a brittle lithosphere: thoughts on the global tectonic styles of the
  {E}arth and {V}enus}.{\BBCQ}
\newblock
\APACjournalVolNumPages{Geophys. J. Int.}{133}{3}{669--682}.
\PrintBackRefs{\CurrentBib}

\bibitem [\protect \citeauthoryear {%
Morris%
\ \BBA {} Canright%
}{%
Morris%
\ \BBA {} Canright%
}{%
{\protect \APACyear {1984}}%
}]{%
morris1984boundary}
\APACinsertmetastar {%
morris1984boundary}%
\begin{APACrefauthors}%
Morris, S.%
\BCBT {}\ \BBA {} Canright, D.%
\end{APACrefauthors}%
\unskip\
\newblock
\APACrefYearMonthDay{1984}{}{}.
\newblock
{\BBOQ}\APACrefatitle {A boundary-layer analysis of {B}\'enard convection in a
  fluid of strongly temperature-dependent viscosity} {A boundary-layer analysis
  of {B}\'enard convection in a fluid of strongly temperature-dependent
  viscosity}.{\BBCQ}
\newblock
\APACjournalVolNumPages{Phys. Earth Planet. Inter.}{36}{3-4}{355--373}.
\PrintBackRefs{\CurrentBib}

\bibitem [\protect \citeauthoryear {%
O'Farrell%
, Lowman%
\BCBL {}\ \BBA {} Bunge%
}{%
O'Farrell%
\ \protect \BOthers {.}}{%
{\protect \APACyear {2013}}%
}]{%
o2013comparison}
\APACinsertmetastar {%
o2013comparison}%
\begin{APACrefauthors}%
O'Farrell, K\BPBI A.%
, Lowman, J\BPBI P.%
\BCBL {}\ \BBA {} Bunge, H\BHBI P.%
\end{APACrefauthors}%
\unskip\
\newblock
\APACrefYearMonthDay{2013}{}{}.
\newblock
{\BBOQ}\APACrefatitle {Comparison of spherical-shell and plane-layer mantle
  convection thermal structure in viscously stratified models with mixed-mode
  heating: Implications for the incorporation of temperature-dependent
  parameters} {Comparison of spherical-shell and plane-layer mantle convection
  thermal structure in viscously stratified models with mixed-mode heating:
  Implications for the incorporation of temperature-dependent
  parameters}.{\BBCQ}
\newblock
\APACjournalVolNumPages{Geophys. J. Int.}{192}{2}{456--472}.
\PrintBackRefs{\CurrentBib}

\bibitem [\protect \citeauthoryear {%
Parmentier%
\ \BBA {} Morgan%
}{%
Parmentier%
\ \BBA {} Morgan%
}{%
{\protect \APACyear {1982}}%
}]{%
parmentier1982thermal}
\APACinsertmetastar {%
parmentier1982thermal}%
\begin{APACrefauthors}%
Parmentier, E.%
\BCBT {}\ \BBA {} Morgan, J.%
\end{APACrefauthors}%
\unskip\
\newblock
\APACrefYearMonthDay{1982}{}{}.
\newblock
{\BBOQ}\APACrefatitle {Thermal convection in non-{N}ewtonian fluids: Volumetric
  heating and boundary layer scaling} {Thermal convection in non-{N}ewtonian
  fluids: Volumetric heating and boundary layer scaling}.{\BBCQ}
\newblock
\APACjournalVolNumPages{J. Geophys. Res.}{87}{B9}{7757--7762}.
\PrintBackRefs{\CurrentBib}

\bibitem [\protect \citeauthoryear {%
Parmentier%
\ \BBA {} Sotin%
}{%
Parmentier%
\ \BBA {} Sotin%
}{%
{\protect \APACyear {2000}}%
}]{%
parmentier2000three}
\APACinsertmetastar {%
parmentier2000three}%
\begin{APACrefauthors}%
Parmentier, E.%
\BCBT {}\ \BBA {} Sotin, C.%
\end{APACrefauthors}%
\unskip\
\newblock
\APACrefYearMonthDay{2000}{}{}.
\newblock
{\BBOQ}\APACrefatitle {Three-dimensional numerical experiments on thermal
  convection in a very viscous fluid: Implications for the dynamics of a
  thermal boundary layer at high {R}ayleigh number} {Three-dimensional
  numerical experiments on thermal convection in a very viscous fluid:
  Implications for the dynamics of a thermal boundary layer at high {R}ayleigh
  number}.{\BBCQ}
\newblock
\APACjournalVolNumPages{Phys. Fluids}{12}{3}{609--617}.
\PrintBackRefs{\CurrentBib}

\bibitem [\protect \citeauthoryear {%
Parmentier%
, Turcotte%
\BCBL {}\ \BBA {} Torrance%
}{%
Parmentier%
\ \protect \BOthers {.}}{%
{\protect \APACyear {1976}}%
}]{%
parmentier1976studies}
\APACinsertmetastar {%
parmentier1976studies}%
\begin{APACrefauthors}%
Parmentier, E.%
, Turcotte, D.%
\BCBL {}\ \BBA {} Torrance, K.%
\end{APACrefauthors}%
\unskip\
\newblock
\APACrefYearMonthDay{1976}{}{}.
\newblock
{\BBOQ}\APACrefatitle {Studies of finite amplitude non-{N}ewtonian thermal
  convection with application to convection in the {E}arth's mantle} {Studies
  of finite amplitude non-{N}ewtonian thermal convection with application to
  convection in the {E}arth's mantle}.{\BBCQ}
\newblock
\APACjournalVolNumPages{J. Geophys. Res.}{81}{11}{1839--1846}.
\PrintBackRefs{\CurrentBib}

\bibitem [\protect \citeauthoryear {%
Puster%
, Jordan%
\BCBL {}\ \BBA {} Hager%
}{%
Puster%
\ \protect \BOthers {.}}{%
{\protect \APACyear {1995}}%
}]{%
puster1995characterization}
\APACinsertmetastar {%
puster1995characterization}%
\begin{APACrefauthors}%
Puster, P.%
, Jordan, T\BPBI H.%
\BCBL {}\ \BBA {} Hager, B\BPBI H.%
\end{APACrefauthors}%
\unskip\
\newblock
\APACrefYearMonthDay{1995}{}{}.
\newblock
{\BBOQ}\APACrefatitle {Characterization of mantle convection experiments using
  two-point correlation functions} {Characterization of mantle convection
  experiments using two-point correlation functions}.{\BBCQ}
\newblock
\APACjournalVolNumPages{J. Geophys. Res.}{100}{B4}{6351--6365}.
\PrintBackRefs{\CurrentBib}

\bibitem [\protect \citeauthoryear {%
Solomatov%
}{%
Solomatov%
}{%
{\protect \APACyear {1995}}%
}]{%
solomatov1995scaling}
\APACinsertmetastar {%
solomatov1995scaling}%
\begin{APACrefauthors}%
Solomatov, V\BPBI S.%
\end{APACrefauthors}%
\unskip\
\newblock
\APACrefYearMonthDay{1995}{}{}.
\newblock
{\BBOQ}\APACrefatitle {Scaling of temperature-and stress-dependent viscosity
  convection} {Scaling of temperature-and stress-dependent viscosity
  convection}.{\BBCQ}
\newblock
\APACjournalVolNumPages{Phys. Fluids}{7}{2}{266--274}.
\PrintBackRefs{\CurrentBib}

\bibitem [\protect \citeauthoryear {%
Solomatov%
\ \BBA {} Moresi%
}{%
Solomatov%
\ \BBA {} Moresi%
}{%
{\protect \APACyear {2000}}%
}]{%
solomatov2000scaling}
\APACinsertmetastar {%
solomatov2000scaling}%
\begin{APACrefauthors}%
Solomatov, V\BPBI S.%
\BCBT {}\ \BBA {} Moresi, L\BHBI N.%
\end{APACrefauthors}%
\unskip\
\newblock
\APACrefYearMonthDay{2000}{}{}.
\newblock
{\BBOQ}\APACrefatitle {Scaling of time-dependent stagnant lid convection:
  Application to small-scale convection on {E}arth and other terrestrial
  planets} {Scaling of time-dependent stagnant lid convection: Application to
  small-scale convection on {E}arth and other terrestrial planets}.{\BBCQ}
\newblock
\APACjournalVolNumPages{J. Geophys. Res.}{105}{B9}{21795--21817}.
\PrintBackRefs{\CurrentBib}

\bibitem [\protect \citeauthoryear {%
Sotin%
\ \BBA {} Labrosse%
}{%
Sotin%
\ \BBA {} Labrosse%
}{%
{\protect \APACyear {1999}}%
}]{%
sotin1999three}
\APACinsertmetastar {%
sotin1999three}%
\begin{APACrefauthors}%
Sotin, C.%
\BCBT {}\ \BBA {} Labrosse, S.%
\end{APACrefauthors}%
\unskip\
\newblock
\APACrefYearMonthDay{1999}{}{}.
\newblock
{\BBOQ}\APACrefatitle {Three-dimensional thermal convection in an isoviscous,
  infinite {P}randtl number fluid heated from within and from below:
  Applications to the transfer of heat through planetary mantles}
  {Three-dimensional thermal convection in an isoviscous, infinite {P}randtl
  number fluid heated from within and from below: Applications to the transfer
  of heat through planetary mantles}.{\BBCQ}
\newblock
\APACjournalVolNumPages{Phys. Earth Planet. Inter.}{112}{3-4}{171--190}.
\PrintBackRefs{\CurrentBib}

\bibitem [\protect \citeauthoryear {%
Stevenson%
, Spohn%
\BCBL {}\ \BBA {} Schubert%
}{%
Stevenson%
\ \protect \BOthers {.}}{%
{\protect \APACyear {1983}}%
}]{%
stevenson1983magnetism}
\APACinsertmetastar {%
stevenson1983magnetism}%
\begin{APACrefauthors}%
Stevenson, D\BPBI J.%
, Spohn, T.%
\BCBL {}\ \BBA {} Schubert, G.%
\end{APACrefauthors}%
\unskip\
\newblock
\APACrefYearMonthDay{1983}{}{}.
\newblock
{\BBOQ}\APACrefatitle {Magnetism and thermal evolution of the terrestrial
  planets} {Magnetism and thermal evolution of the terrestrial planets}.{\BBCQ}
\newblock
\APACjournalVolNumPages{Icarus}{54}{3}{466--489}.
\PrintBackRefs{\CurrentBib}

\bibitem [\protect \citeauthoryear {%
Tackley%
}{%
Tackley%
}{%
{\protect \APACyear {1996}}%
}]{%
tackley1996ability}
\APACinsertmetastar {%
tackley1996ability}%
\begin{APACrefauthors}%
Tackley, P\BPBI J.%
\end{APACrefauthors}%
\unskip\
\newblock
\APACrefYearMonthDay{1996}{}{}.
\newblock
{\BBOQ}\APACrefatitle {On the ability of phase transitions and viscosity
  layering to induce long wavelength heterogeneity in the mantle} {On the
  ability of phase transitions and viscosity layering to induce long wavelength
  heterogeneity in the mantle}.{\BBCQ}
\newblock
\APACjournalVolNumPages{Geophys. Res. Lett.}{23}{15}{1985--1988}.
\PrintBackRefs{\CurrentBib}

\bibitem [\protect \citeauthoryear {%
Travis%
\ \BBA {} Olson%
}{%
Travis%
\ \BBA {} Olson%
}{%
{\protect \APACyear {1994}}%
}]{%
travis1994convection}
\APACinsertmetastar {%
travis1994convection}%
\begin{APACrefauthors}%
Travis, B.%
\BCBT {}\ \BBA {} Olson, P.%
\end{APACrefauthors}%
\unskip\
\newblock
\APACrefYearMonthDay{1994}{}{}.
\newblock
{\BBOQ}\APACrefatitle {Convection with internal heat sources and thermal
  turbulence in the {E}arth's mantle} {Convection with internal heat sources
  and thermal turbulence in the {E}arth's mantle}.{\BBCQ}
\newblock
\APACjournalVolNumPages{Geophys. J. Int.}{118}{1}{1--19}.
\PrintBackRefs{\CurrentBib}

\bibitem [\protect \citeauthoryear {%
Turcotte%
\ \BBA {} Oxburgh%
}{%
Turcotte%
\ \BBA {} Oxburgh%
}{%
{\protect \APACyear {1967}}%
}]{%
turcotte1967finite}
\APACinsertmetastar {%
turcotte1967finite}%
\begin{APACrefauthors}%
Turcotte, D.%
\BCBT {}\ \BBA {} Oxburgh, E.%
\end{APACrefauthors}%
\unskip\
\newblock
\APACrefYearMonthDay{1967}{}{}.
\newblock
{\BBOQ}\APACrefatitle {Finite amplitude convective cells and continental drift}
  {Finite amplitude convective cells and continental drift}.{\BBCQ}
\newblock
\APACjournalVolNumPages{J. Fluid Mech.}{28}{1}{29--42}.
\PrintBackRefs{\CurrentBib}

\bibitem [\protect \citeauthoryear {%
Vilella%
\ \BBA {} Deschamps%
}{%
Vilella%
\ \BBA {} Deschamps%
}{%
{\protect \APACyear {2018}}%
}]{%
vilella2018temperature}
\APACinsertmetastar {%
vilella2018temperature}%
\begin{APACrefauthors}%
Vilella, K.%
\BCBT {}\ \BBA {} Deschamps, F.%
\end{APACrefauthors}%
\unskip\
\newblock
\APACrefYearMonthDay{2018}{}{}.
\newblock
{\BBOQ}\APACrefatitle {Temperature and heat flux scaling laws for isoviscous,
  infinite {P}randtl number mixed heating convection} {Temperature and heat
  flux scaling laws for isoviscous, infinite {P}randtl number mixed heating
  convection}.{\BBCQ}
\newblock
\APACjournalVolNumPages{Geophys. J. Int.}{214}{1}{265--281}.
\PrintBackRefs{\CurrentBib}

\bibitem [\protect \citeauthoryear {%
Vilella%
\ \BBA {} Kaminski%
}{%
Vilella%
\ \BBA {} Kaminski%
}{%
{\protect \APACyear {2017}}%
}]{%
vilella2017fully}
\APACinsertmetastar {%
vilella2017fully}%
\begin{APACrefauthors}%
Vilella, K.%
\BCBT {}\ \BBA {} Kaminski, E.%
\end{APACrefauthors}%
\unskip\
\newblock
\APACrefYearMonthDay{2017}{}{}.
\newblock
{\BBOQ}\APACrefatitle {Fully determined scaling laws for volumetrically heated
  convective systems, a tool for assessing habitability of exoplanets} {Fully
  determined scaling laws for volumetrically heated convective systems, a tool
  for assessing habitability of exoplanets}.{\BBCQ}
\newblock
\APACjournalVolNumPages{Phys. Earth Planet. Inter.}{266}{}{18--28}.
\PrintBackRefs{\CurrentBib}

\bibitem [\protect \citeauthoryear {%
Weller%
, Lenardic%
\BCBL {}\ \BBA {} Moore%
}{%
Weller%
\ \protect \BOthers {.}}{%
{\protect \APACyear {2016}}%
}]{%
weller2016scaling}
\APACinsertmetastar {%
weller2016scaling}%
\begin{APACrefauthors}%
Weller, M\BPBI B.%
, Lenardic, A.%
\BCBL {}\ \BBA {} Moore, W\BPBI B.%
\end{APACrefauthors}%
\unskip\
\newblock
\APACrefYearMonthDay{2016}{}{}.
\newblock
{\BBOQ}\APACrefatitle {Scaling relationships and physics for mixed heating
  convection in planetary interiors: Isoviscous spherical shells} {Scaling
  relationships and physics for mixed heating convection in planetary
  interiors: Isoviscous spherical shells}.{\BBCQ}
\newblock
\APACjournalVolNumPages{J. Geophys. Res. Solid Earth}{121}{10}{7598--7617}.
\PrintBackRefs{\CurrentBib}

\end{thebibliography}

%
%
%
%
%

\end{document}